\documentclass[aps,prd,preprint,superscriptaddress,showkeys]{revtex4}
\usepackage{graphicx}

\begin{document}

%\preprint{}

\title{Thermoelastic dissipation in inhomogeneous media: loss
measurements and
displacement noise in coated test masses for interferometric
gravitational wave detectors}


\author{M.M. Fejer$^{1}$, S. Rowan$^{1,2}$, G. Cagnoli$^{2}$, D.R.M.
Crooks$^{2}$, A. Gretarsson$^{3}$, G.M. Harry$^{4}$, J. Hough$^{2}$,
S.D. Penn$^{5}$, P.H.
Sneddon$^{2}$, and S.P. Vyatchanin,$^{6}$}
%\thanks{}
\affiliation{$^{1}$ Edward L. Ginzton Laboratory, Stanford University,
Stanford, CA 94305\\
$^{2}$ IGR, Dept. of Physics and Astronomy, University of Glasgow,
Glasgow G12 8QQ\\
$^{3}$ LIGO Livingston Laboratory, 19100 LIGO Lane, Livingston, LA
70754 \\
$^{4}$ LIGO Laboratory, Massachusetts Institute of Technology, \\Cambridge,
MA 02139\\
$^{5}$ Dept. of Physics, Hobart and William Smith Colleges, Geneva, NY
14456 \\
$^{6}$~Physics Faculty, Moscow State University, Moscow 119899, Russia}


\date{\today}

\begin{abstract}
The displacement noise in the test mass mirrors of interferometric
gravitational wave detectors is proportional to their elastic
dissipation at the observation frequencies. In this paper, we
analyze one fundamental source of dissipation in thin coatings,
thermoelastic damping associated with the dissimilar thermal and
elastic properties of the film and the substrate. We obtain
expressions for the thermoelastic dissipation factor necessary to
interpret resonant loss measurements, and for the spectral density
of displacement noise imposed on a Gaussian beam reflected from
the face of a coated mass. The predicted size of these effects is
large enough to affect the interpretation of loss measurements,
and to influence design choices in advanced gravitational wave
detectors.
\end{abstract}


\maketitle

\section{Introduction}
\label{sec:intro}
While recent results indicate that the elastic losses available in
bulk materials such as silica and sapphire are adequate to achieve
the design goals for displacement noise in next generation
interferometric gravitational-wave detectors, the losses
associated with the multilayer dielectric mirrors deposited on the
faces of the mass are large enough to contribute significantly to
the total noise of the system. The origin of these coating losses
is not yet clear. In this paper, we investigate an intrinsic
dissipation mechanism, thermoelastic effects associated with a
thin film on a bulk substrate.
Thermoelastic dissipation has been known since Zener's work in the
1930's.~\cite{Zener} In homogeneous solids, it is associated with the
irreversible flow of heat driven by temperature gradients associated
with strain gradients in the solid. These effects lead to the well
known result for the damping of flexural vibrations in a thin beam,
where the heat flows from the side of the beam in compression to the
side in tension.\cite{LifshitzandR} The maximum dissipation
$\phi_{\mathrm{max}}$ in this case is a function only of the material
properties and not the beam dimensions, $\phi_{\mathrm{max}} = Q^{-1} \sim
E\alpha^{2}T/C$, where $E$ is the Young's modulus, $\alpha$ is the
thermal expansion coefficient, $C$ is the volumetric heat capacity,
and $T$ is the background temperature. The dissipation peak occurs at
a frequency $\omega_{p} \sim 1/\tau$ where $\tau \sim l^{2}C/k$
is the thermal diffusion time through a beam of thickness $l$ given
the thermal conductivity $k$.

More recently, thermoelastic dissipation in homogeneous test masses
has been analyzed as a source of noise in gravitational
wave detectors,~\cite{Braginsky} where the heat can be viewed as diffusing
radially in the compressed region associated with a
Gaussian-distributed pressure field on the surface of the mass, as is
used in Levin's method for analyzing displacement noise.\cite{Levin}
Characteristic of both these examples is the presence of a nonuniform
strain field, necessary to create a temperature gradient to drive the
thermal diffusion in a homogeneous medium.

In an inhomogeneous body, temperature gradients can be generated
in a {\it uniform} strain field, so that thermoelastic dissipation
can be expected even in the absence of stress or strain gradients.
For the case of a coated test mass, if the thermoelastic
properties of the film are different from those in the substrate,
we can anticipate that thermal diffusion and hence thermoelastic
dissipation will occur. An estimate of the size of the effect can
be obtained by comparison with the flexural damping of a thin
beam. Replacing the thermal expansion coefficient by the
difference between these coefficients in the film and substrate,
and assuming for simplicity that the other pertinent material
parameters are the same, we have $\phi_{\rm{max}} \sim
E(\alpha_{f}-\alpha_{s})^{2}T/C$, and again expect the peak
response to occur for frequencies corresponding to the thermal
diffusion time through the film. If we consider a film with the
thermal expansion coefficient of alumina on a substrate with the
thermal expansion of silica, and take the other parameters to be
those of silica we find $\phi_{\mathrm{max}} \sim 3 \times
10^{-4}$, comparable to the elastic losses measured in optical
coatings.\cite{Crooks, Harry} For a 5-micron-thick film with these
properties the dissipation maximum occurs at $\sim 5
\mathrm{\,kHz}$, corresponding to the thermal diffusion time
through the film. This frequency is in the range typically sampled
by mechanical loss measurements, and not far from the frequency
band of interest for gravitational wave detection.

It appears that a more quantitative investigation of these effects is
necessary to evaluate their implications for characterization of test
masses as well as for gravitational wave detectors themselves. We
consider in this paper two questions associated with the
thermoelastic mechanism: what is the effective dissipation in the
situation characteristic of resonator measurements of elastic loss, and
what is the expected spectral density of displacement noise in the
situation characteristic of a test mass in a gravitational wave
detector.

The key results from the analyses contained here are: (a) the
derivation of an expression for $\phi_{\mathrm{tot}, \,
\parallel}$, the thermoelastic dissipation expected in a coated
test mass undergoing deformations of the type expected in
mechanical loss measurements and (b) the derivation of an
expression for $S_{x}(f)$, the power spectral density of
thermoelastic displacement noise associated with the dielectric
mirror coating on a test-mass substrate.

A recent independent calculation of the spectral density of thermal noise
in the low frequency limit agrees, for the case where the thermoelastic
properties of the film and substrate other than the thermal expansion
coefficient are identical, with the results given here taken to that same
limit.~\cite{Braginskyfilm, Brag2}. 

In reference~\cite{Brag2}, the difference between the expressions for 
thermoelastic thermal noise presented here and in~\cite{Braginskyfilm} for cases  
where the elastic properties of the film and substrate differ is noted. After 
discussions with the authors this difference has been resolved in 
favor of the results presented here.

\section{Sketch of the calculation} \label{sec:sketch}

In this section, we sketch the calculation of the thermoelastic
dissipation and the displacement noise in an inhomogeneous medium.
Details of the calculation are given in section~\ref{sec:DetailedCalc} and
appendices.

There are three steps to calculating the thermoelastic loss in
the
coating:
\begin{enumerate}
       \item Obtain the oscillatory thermal field associated with the
zeroth-order elastic fields,
       \item Calculate the complex first-order
elastic fields generated by the spatially varying oscillatory thermal
field, and then
       \item Calculate the power dissipated by the interaction
of the zeroth- and first-order elastic fields.
\end{enumerate}
Throughout we will consider only linear
thermoelasticity, retaining terms up to first order in the
oscillatory thermal field. Therefore the stored energy can be taken to be
proportional to the square of zeroth-order elastic fields, while the
imaginary
part of the product of the zeroth-order elastic fields and the elastic
fields induced
by the thermal wave represent the relevant average dissipated power. Zeroth
and first order quantities are denoted by subscripts 0, 1,
respectively.

The geometry we consider consists of a film of thickness $l$ and
thermal expansion coefficient $\alpha_f$ on a substrate with $ \alpha
_s $ whose thermophysical properties are all possibly different from
those of the film. We take the surface normal to be in the $-z$
direction,
and the surface located at $z=0$, so that the film extends from $z=0$
to $z=l$, and the substrate from $z=l$ to $z=\infty$.

To simplify the analysis, we assume that the multilayer film can be
approximated as a uniform film with appropriately averaged
properties, and assume that the film is thin enough and the thermal
diffusion length at the frequencies of interest short enough compared
to any relevant transverse dimension (e.g. the dimensions of the
object itself, or the radius of the Gaussian
beam interrogating its surface) that only the
thermal diffusion normal to the surface of the mass need be
considered. Since the thermal diffusion lengths for frequencies of
interest are on the order of or longer than the total film thickness,
the description of the film in terms of its average properties
appears reasonable, but we also consider the case of a film whose
thermal expansion (but no other parameters) varies periodically
through its thickness, as a simple model to explore any unexpected
effects that might arise from the neglected microstructure of the
film.

The point of departure for the calculation is the thermal diffusion
equation, driven by a thermoelastic source term, which for the
assumed one-dimensional heat flow can be cast in the
form~\cite{LifshitzandR}
\begin{equation}\label{eq:1}
i\omega \theta _j  - \kappa _j
\frac{{d^2 \theta _j }}{{d z^2 }} =  - \frac{{E_j
\alpha _j T}}{{(1 - 2\nu _j )C_j }}i\omega \sum\limits_{i = 1}^3
{\varepsilon _{0,ii,j} }
\end{equation}
where $\theta_j(z)$ is the time-varying temperature with $\exp(i
\omega t)$ time dependence assumed, $\kappa_j=k_{j}/C_j$ is the
thermal diffusivity, $E_j$ is the
Young's modulus, $\nu_j$ is the Poisson ratio, $T$ is the background
temperature, $C_j$ is the heat capacity per unit volume,
$\varepsilon_{0,ii,j}$ is the zeroth-order $i$-polarized
compressional strain, and $j=f,s$ indicates quantities evaluated in
the film and the substrate, respectively.

To obtain the source term, we need the zeroth-order compressional
strains. Different combinations of zeroth-order elastic fields are
relevant in different situations. We will assume that the
transverse variation of the zeroth-order elastic fields is slow
compared to the thickness of the film, so their only possible
variation is in the $z$-direction, and that that variation results
only from the possible discontinuity of the elastic properties at
the film-substrate boundary. Note that this statement regarding
the {\it z}-dependence applies only to the {\it {zeroth}}-order
elastic fields; as we will see, the thermal fields and the
first-order elastic fields they generate have a {\it z}-dependence
that arises from the propagation of the oscillatory thermal wave
itself. Under these assumptions, we can specify the zeroth-order
fields in terms of three quantities that, due to the elastic
boundary conditions, do not vary over the length scales relevant
to this problem: the axial stress $\sigma_{0} \equiv
\sigma_{0,zz}$, the symmetric combination of in-plane strains (the
dilation) $\varepsilon_{0} \equiv
(\varepsilon_{0,xx}+\varepsilon_{0,yy})/2$, and the antisymmetric
combination of in-plane strains
$\varepsilon_{0,xx}-\varepsilon_{0,yy}$. All of the other
components of the zeroth-order elastic fields can be derived from
these three, as is established in Appendix \ref{ap:zeroth}. The
anti-symmetric combination of strains, which is a pure shear along
axes rotated $\pi/4$ to $x$ and $y$, does not interact
thermoelastically, and can be neglected in the remainder of the
analysis, as is established in more detail in section
\ref{sec:unifilm}.

Given these zeroth-order elastic fields, we can evaluate the source
term in Eq. \ref{eq:1} and solve for the oscillatory thermal wave,
$\theta(z)$, as discussed in section \ref{sec:Thermal fields} and
Appendix \ref{ap:theta}. This thermal wave, in turn, generates a
first-order elastic field, with compressional components
$\varepsilon_{1,ii}$ and $\sigma_{1,ii}$. The thermoelastic coupling
enters into the formulation through the elastic
equilibrium equations and modified Hooke's law, which can be cast for the
one-dimensional case
considered here from Eqs. 7.8 and 6.2 of~\cite{Landau},
\begin{equation}\label{eq:2a}
\frac{d}{{dz}}\left[ {\varepsilon _{1,xx,j}  + \varepsilon _{1,yy,j}  +
2(1 -
\nu_j )\varepsilon _{1,zz,j}  - 2(1 + \nu_j )\alpha \theta_j }
\right] = 0
\end{equation} and
\begin{equation}\label{eq:2b}
\sigma _{1,ii,j}  = \frac{E_j}{{1 + \nu_j }}\left[ {\varepsilon _{1,ii,j}
+
\frac{\nu_j }{{1 - 2\nu_j }}(\varepsilon _{1,xx,j}  + \varepsilon
_{1,yy,j}  +
\varepsilon _{1,zz,j} )} \right] - \frac{{E_j\alpha_j \theta_j }}{{1 -
2\nu_j }}
\,\, ,
\end{equation}
where $j = f,s$ represents fields and material properties in the film and
substrate, respectively. With the thermal fields obtained by solving Eq.
\ref{eq:1}, we can
obtain from Eqs.~\ref{eq:2a} and~\ref{eq:2b} the first-order elastic
fields, as given in Appendix
\ref{ap:first}.

The rate at which work is done per unit volume in a deformed body is in
general given by
\[
p = \sigma _{ik} \frac{{d\varepsilon _{ik} }}{{dt}} \nonumber \\
\]
and the dissipated power density by
\begin{equation} \label{eq:3}
    p_{\mathrm{diss}}\approx -\frac{\omega }{2}\sum\limits_{i = 1}^3
{\mbox{Im}\left[{
\sigma _{0,ii}^* \varepsilon _{1,ii}  + \sigma _{1,ii}^* \varepsilon
_{0,ii}^{} }\right]}.
\end{equation}
where the second form is specified to our problem with sinusoidal
fields and only longitudinal strains, and takes into account that the
zeroth-order fields are real, so that dissipation first appears in the
product of first- and zeroth-order fields. Integrating the dissipated
power density over $0 \leq z <
\infty$, we obtain the dissipated power per unit area, given in
section \ref{sec:EnergyDensity}. While this is the essential quantity
of interest, it is convenient for comparison to experimental
measurements of elastic Q to define an effective dissipation factor,
$\phi$. Since the thermoelastic dissipation is nonlocal, the choice
of stored energy with which to make such a definition is somewhat
arbitrary. A reasonable choice, and the one that results in a value for
$\phi$ directly comparable to that derived from experimental results, is
the elastic energy stored in
the film. With this choice, as described in Section
\ref{sec:unifilm}, we find that for an elastic field with specified
in-plane strain and vanishing axial stress, as would be appropriate
for a measurement of the elastic Q of a mass coated with a uniform film
of thickness $\it l$, the total thermoelastic loss, $\phi_{tot}$ (Eq.
\ref{eq:phimod2}) is well approximated by
\begin{eqnarray} \label{eq:4}
\phi_{l, \, \parallel}=\frac{2 E_f \alpha_f^2
T}{C_f(1-\nu_f)}\left[ 1 - \frac{\alpha_s}{\overline{\alpha}_f}
\frac{E_s (1-\nu_f)}{E_f (1-\nu_s)}\frac{C_f}{C_s} \right]^2
g(\omega)
\end{eqnarray}
where the frequency dependence is contained in the function
$g(\omega)$ defined by
\begin{equation}  \label{eq:5}
g(\omega)\equiv {\mathop{\rm Im}\nolimits} \left[ - \frac
{1}{\sqrt{i\omega\tau_f}} \frac{\sinh
(\sqrt{i\omega\tau_f})}{\cosh(\sqrt{i\omega\tau_f})+R\sinh(\sqrt{i\omega\tau_f})}\right]
\,\, ,
\end{equation}
where $\omega = 2\pi f, \tau_f\equiv l^2/\kappa_f=l^2C_f/k_{f}$ is the thermal
diffusion time across the film, and $R \equiv \sqrt{k_f C_f/k_s
C_s}$, with $k_j$ and $C_j$ the thermal conductivity and volumetric
heat capacity, respectively. Eqs. \ref{eq:4} and \ref{eq:5}
(or the form for a multilayer in Eq. \ref{eq:phiunisum}) constitute one of
the two key results of this paper. The frequency
dependence represented by $g(\omega)$ is discussed at length in
section \ref{sec:FrequencyDependence}. The quantitative implications
for measurements of thermoelastic dissipation in several material
systems are discussed in section \ref{subsec:numQ}.

These numerical results indicate that thermoelastic losses
associated with the coating are comparable to those obtained in
experimental measurements of elastic loss, which suggests that
their contribution to the total displacement noise budget for a
test mass could be significant. While one could form an expression
for the thermal noise imposed on a Gaussian beam interrogating a
coated test mass by inserting $\phi_{l, \, \parallel}$ from Eq.
\ref{eq:4}, and a corresponding one for $\phi_{l, \, \perp}$ from
Eq. \ref{eq:phiuniperp} into one of the expressions developed for
thermal noise in coated test masses~\cite{Nakagawa, Harry}, the
result would only be approximate because of the thermoelastic
coupling between in-plane and normal strains. A direct calculation
of the thermal noise can instead be carried out using Levin's
formulation, calculating the power dissipated by the thermoelastic
mechanism when a pressure field with the same radial distribution
as the optical intensity field is applied to the coated mass. This
calculation, using the zeroth-order elastic fields obtained
in~\cite{Harry}, is carried out for arbitrary frequency in section
\ref{sec:ThermalNoise}. We find
\begin{eqnarray} \label{eq:6}
S_x(f) df &=&
\frac{8k_BT^{2}}{\pi^{2}f}\frac{l}{w^2}\frac{\alpha_s^{2}C_f}{C_s^{2}}(1+\nu_s)^{2}\Delta^2
g(\omega) df \nonumber \\
&\to& \frac{8 \sqrt{2}k_B T^2}{\pi
\sqrt{\omega}}\frac{l^2}{w^2}(1+\nu_s)^2 \,
\frac{C_f^2}{C_s^2}\frac{\alpha_s^2}{\sqrt{k_s C_s}} \, \Delta^2  df
\,\, ,
\end{eqnarray}
where $\Delta^{2}$ is a dimensionless positive-definite
combination of material constants that vanishes when the film and
substrate are identical,
\begin{equation} \label{eq:7}
\Delta^{2}\equiv\left\{\frac{C_s}{2 \alpha_s
C_f}\frac{\alpha_f}{(1-\nu_f)}\left[
\frac{1+\nu_f}{1+\nu_s}+(1-2\nu_s)\frac{E_f}{E_s}\right]-1
\right\}^2 \,\, ,
\end{equation}
$g(\omega)$ is the same frequency dependence as defined in Eq.
\ref{eq:5}, and the second form holds for low frequencies obeying
$\omega < 1/\tau_f$. Note that the results in section
\ref{subsec:numthermalnoise} show that the limiting form must be
used in the gravitational-wave-detection band only with caution.
Eq. \ref{eq:6} (or the corresponding Eq.~\ref{eq:Sxavgsum} for a
multilayer) is the other key result of this paper; quantitative
implications for several plausible mass/coating combinations are
presented in Section \ref{subsec:numthermalnoise}.

While the results for $\phi_{l, \, \parallel}$ and $S_x(f)$ in
Eqs.~\ref{eq:4} and \ref{eq:6} are calculated for a uniform film,
most optical coatings of course will consist of a large number of
layers. In such a multilayered coating, there are two
thermoelastic dissipation peaks, one at a frequency related to the
thermal diffusion time through the entire film, and one at a
frequency related to the thermal diffusion time through an
individual layer. These interlayer effects are investigated in
section \ref{sec:modfilm}. It is seen there together with section
\ref{sec:FrequencyDependence} that for problems of interest, the
thermoelastic effects are dominated by contributions from the
thermal diffusion through the film, so that a description of the
multilayer in terms of a set of averaged properties appears
appropriate. The subtleties of the averaging process are
investigated in Appendix \ref{ap:avg}, where it is seen that the
average of various products of material quantities is required in
addition to the average of the quantities themselves. Specializing
to a periodic multilayer of total thickness $l$ with $N$
alternating layers of materials $a$ and $b$ in thicknesses of
$d_a$ and $d_b$, the result for $\phi_{l, \,
\parallel}$ from Eq.~\ref{eq:phiuniavg},
\begin{eqnarray} \label{eq:phiunisum}
\phi_{l, \, \parallel}= \frac{2 C_F T}{\left( \frac{E}{1-\nu}
\right)_{\mathrm{avg}}} \left[  \frac{1}{C_F}\left(\frac{E
\alpha}{1-\nu} \right)_{\mathrm{avg}}-\frac{1}{C_s}\frac{E_s
\alpha_s}{1-\nu_s} \right]^2 g(\omega)
\end{eqnarray}
where the frequency dependence is contained in the same function
$g(\omega)$ as defined in Eq.~\ref{eq:5} with $\tau_f \to \tau_F$ where
$\tau_F = l^2 / \kappa_F$, and the result for $S_x(f)$ from Eq.~\ref{eq:sxlow},
\begin{eqnarray} \label{eq:Sxavgsum}
  S_x(f)
&=&\frac{8k_BT^{2}}{\pi^{2}f}\frac{l}{w^2}\frac{\alpha_s^{2}C_F}{C_s^{2}}(1+\nu_s)^{2}
\tilde{\Delta}^2 g(\omega)\nonumber\\
& \to& \frac{8 \sqrt{2}k_B T^2}{\pi
\sqrt{\omega}}\frac{l^2}{w^2}(1+\nu_s)^2 \,
\frac{C_F^2}{C_s^2}\frac{\alpha_s^2}{\sqrt{k_s C_s}} \, \tilde{\Delta}^2
\,\, ,
\end{eqnarray}
and Eq.~\ref{eq:deltilde}
\begin{equation}
\tilde{\Delta}^2\equiv\left\{ \frac{C_s}{2 \alpha_s C_F} \left(
\frac{\alpha}{1-\nu}\left[
\frac{1+\nu}{1+\nu_s}+(1-2\nu_s)\frac{E}{E_s}\right]\right)_{\mathrm{avg}}-1
\right\}^2
\,\, ,
\end{equation}
can be stated in terms of an averaging operator defined in Eq.
\ref{eq:avgdef} as
\begin{equation}
(X)_{\mathrm{avg}}\equiv \frac{d_a}{d_a+d_b} X_a + \frac{d_b}{d_a+d_b} X_b
\,\, ,
\end{equation}
and volume-averaged material properties $C_F$ and $\kappa_F$ defined in
Eq. \ref{eq:thermavg}. The second form of Eq.~\ref{eq:Sxavgsum} is, like 
the second form of Eq.~\ref{eq:6}, a low frequency limit valid for$\omega < 
1/\tau_f$.

\section{Numerical results}
\subsection{Numerical results for dissipation in Q measurements}
\label{subsec:numQ}
The mechanical loss factors (or equivalently Q factors) of dielectric
coatings
applied to test mass substrates may be obtained
experimentally.
In a typical measurement of this type, a subset of the
vibrational resonant modes of
a coated substrate are individually excited above some background level
and the decay of the amplitude of vibration of the face of
the sample measured as a function of
time. From this measurement, and a model of the distribution of the
stored energy in each mode of a coated sample, the mechanical loss
factors of the dielectric coating at each of the resonant frequencies
of the sample may be obtained.

As part of the loss measurement process described above, the
coated samples experience periodic strains. If there exists a
difference in the thermoelastic properties of the dielectric
coating and the substrate, then as shown in this paper, there will
be thermoelastic dissipation. Eq.~\ref{eq:phiunisum} can be used
to calculate the thermoelastic dissipation in a coating both at
the frequencies typical of mechanical loss measurements, and at
frequencies of interest for gravitational-wave detection.

The expected thermoelastic
loss associated with a coating on a substrate is a direct function
of the material parameters for the particular substrates and coatings
chosen. Current interferometric detectors use fused-silica substrates
with coatings formed from alternating layers of SiO$_2$ (refractive
index n = 1.45) and
Ta$_{2}$O$_{5}$ (refractive index n = 2.03), each layer being of
$\lambda/4$ optical thickness, with
$\lambda$ = 1064nm.
The mirrors in future upgrades to current detectors are expected to
have substrates of either fused silica or
sapphire.  The choice of appropriate mirror
coating materials
is a subject of ongoing study~\cite{Crooks, Harry, Penn}, with the two
coatings currently under most intense study being alternating layers of
SiO$_2$ and
Ta$_2$O$_5$, or Al$_2$O$_3$ (refractive index n = 1.63) and Ta$_2$O$_5$.

To estimate the expected level of thermo-elastic loss for the
mirror/substrate coating combinations above,
Eq.~\ref{eq:phiunisum} was used. In each case a coating thickness
equivalent to thirty alternating quarter-lambda layers of the
coating materials was chosen. The numerical values used for the
properties of the mirror substrates are shown in the tables below.

\begin{table}[h]
\caption{Fused silica substrate
properties~\cite{EOHandbook}~\cite{Musikant}}
\begin{ruledtabular}
\begin{tabular}{rrrrrrlllll}
&&&&&$\alpha_s$   &=  5.1 x 10$^{-7}$ K$^{-1}$ &&&&\\
&&&&&$E_s$        &= 7.2 x  10$^{10}$ Nm$^{-2}$ &&&&  \\
&&&&&$C_{s}$  &=   746 Jkg$^{-1}$K$^{-1}$ x 2200 kgm$^{-3}$&&&&\\
&&&&&    &= 1.64 x 10$^{6}$ JK$^{-1}$m$^{-3}$ &&&& \\
&&&&&$k_{s}$     &=  1.38 Wm$^{-1}$ K$^{-1}$ &&&&\\
&&&&&$\nu_{s}$    &= 0.17 &&\\
\end{tabular}
\end{ruledtabular}
\end{table}

\begin{table}[h]
\caption{Sapphire substrate
properties~\cite{EOHandbook}~\cite{Musikant}~\cite{Touloukian}}
\begin{ruledtabular}
\begin{tabular}{rrrrrrlllll}
&&&&&$\alpha_s$   &=  5.4 x 10$^{-6}$ K$^{-1}$&&&&\\
&&&&&$E_s$    &=  4 x 10$^{11}$ Nm$^{-2}$ &&&&\\
&&&&&$C_{s}$  &=  777 Jkg$^{-1}$K$^{-1}$ x 3980 kgm$^{-3}$ &&&&\\
&&&&&    &=  3.09 x 10$^{6}$ JK$^{-1}$m$^{-3}$ &&&&\\
&&&&&$k_{s}$     &= 33 Wm$^{-1}$ K$^{-1}$&&&&\\
&&&&&$\nu_{s}$     &= 0.23&&&& \\
\end{tabular}
\end{ruledtabular}
\end{table}

Choosing appropriate material parameters for the multi-layer
ion-beam-sputtered dielectric coatings is made more difficult by
the fact that thermo-physical properties of these types of
coatings are not well characterized. Absent better information,
the properties of the amorphous SiO$_2$ and Al$_2$O$_3$ present in
the films were assumed to be the same as the bulk values for
amorphous fused silica and crystalline sapphire. The numerical
values used for the properties of Ta$_2$O$_5$ are summarized in
table~\ref{tab:tantalafilm}.

\begin{figure}[!t]
\includegraphics[width=8.9cm]{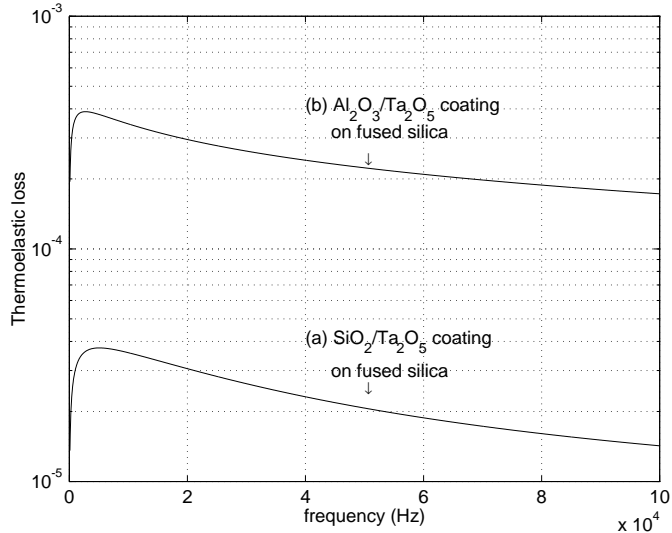}
\caption{Calculated frequency dependence of the thermoelastic
losses of fused silica substrates with dielectric multilayer
coatings formed from thirty alternating quarter-lambda layers of
(a) SiO$_2$ and Ta$_{2}$O$_{5}$ or (b) Al$_2$O$_3$ and
Ta$_{2}$O$_{5}$. Shown are loss factors for frequencies up to 100
kHz.} \label{fig:newforcompexp1}
\end{figure}

\begin{figure}[!h]
\includegraphics[width=8.6cm]{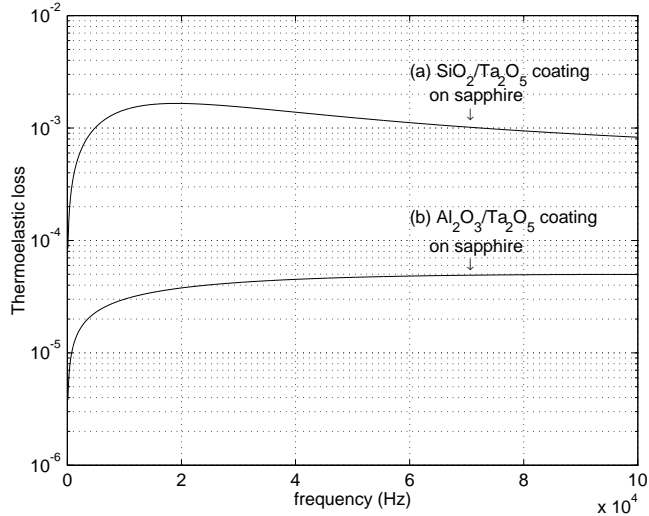}
\caption{Frequency dependence of the thermoelastic losses of
sapphire substrates with dielectric multilayer coatings formed
from thirty alternating quarter-lambda layers of (a) SiO$_2$ and
Ta$_{2}$O$_{5}$ or (b) Al$_2$O$_3$ and Ta$_{2}$O$_{5}$. Shown are
loss factors for frequencies up to 100 kHz.}
\label{fig:newforcompexp2}
\end{figure}

\begin{table}[h!]
\caption{Properties used for Ta$_2$O$_5$
in thin film form.}\label{tab:tantalafilm}
\begin{ruledtabular}
\begin{tabular}{rrrrrrlllll}
&&&&& $\alpha$      &=   3.6 x 10$^{-6}$ K$^{-1}$ &&&&\\
&&   && &$E$    &=  1.4 x 10$^{11} $Nm$^{-2}$ &&&&\\
&&    &&&$C$      &=  306 Jkg$^{-1} $K$^{-1}$ x 6850 kgm$^{-3} $&&&&\\
&&   && &         &=  2.1 x 10$^{6}$ JK$^{-1}$m$^{-3}$  &&&&\\
&&  &&  &$k$     &=  33 Wm$^{-1}$K$^{-1}$ &&&&    \\
&&  &&  &$\nu$    &=  0.23&&&&\\
\end{tabular}
\end{ruledtabular}
\end{table}
Values for Young's modulus and
density of Ta$_2$O$_5$ in thin film form were taken from
reference~\cite{MartinY}. The coefficient of thermal expansion for
Ta$_2$O$_5$ film was taken from~\cite{Tien} and a value for the
specific heat capacity of Ta$_2$O$_5$ (bulk)
from~\cite{oxidehandbook}. No values for thermal conductivity or
Poisson's ratio were available for Ta$_2$O$_5$; absent better
information we take these to be closer to those of sapphire than
silica, and assign them the same values as used for Al$_2$O$_3$.
Using Eq.~\ref{eq:phiunisum} the thermoelastic losses from
coatings of either SiO$_2$/Ta$_2$O$_5$ or Al$_2$O$_3$/Ta$_2$O$_5$
applied to silica and sapphire substrates were calculated for
frequencies up to 100 kHz, a typical range of interest for
measurements of mechanical loss. The estimated loss factors are
plotted in figures~\ref{fig:newforcompexp1}
and~\ref{fig:newforcompexp2}.

% entire page. There is no need to do explicit centering.

% \begin{figure}
% \end{figure}

It can be seen from these figures that in general, the calculated
magnitude of the
thermoelastic losses from these mirror and coating combinations can be of
the order
of a few $10^{-5}$ to approximately $10^{-3}$, comparable to the
levels of coating loss factors predicted to be significant in
estimations of the thermal noise level in advanced gravitational wave
detectors~\cite{Crooks, Harry}.

In previous work~\cite{Penn}, measurements have been made
of the mechanical loss in the frequency range from
~2.8kHz to ~73kHz, for dielectric coatings of SiO$_2$/Ta$_{2}$O$_{5}$
applied to fused silica substrates. The measured coating loss
factors were found to be of the order of 2.8 x 10$^{-4}$. It can be
seen from figure~\ref{fig:newforcompexp1}, curve (a) that the
estimated thermoelastic
losses for this
particular coating/substrate combination are much
smaller than the measured losses.  This suggests that the measured losses
are not predominantly thermoelastic in origin, and are associated
with some other form of dissipation.

\begin{figure}[!ht!]
\includegraphics[width=9.5cm]{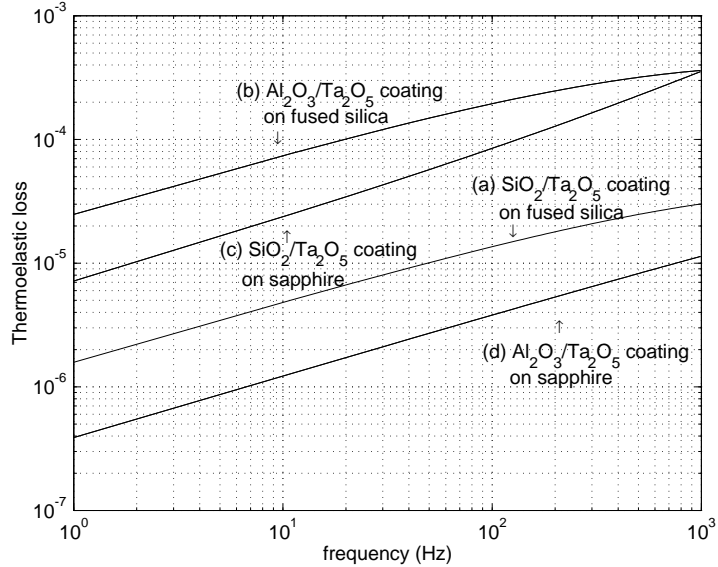}
\caption{Calculated frequency dependence of the thermoelastic losses of (a) an
SiO$_2$/Ta$_{2}$O$_{5}$ coating applied
to a silica substrate (b) an Al$_2$O$_3$/Ta$_{2}$O$_{5}$ applied
to a silica substrate (c) an
SiO$_2$/Ta$_{2}$O$_{5}$ coating applied
to a sapphire substrate and (d) an
Al$_2$O$_3$/Ta$_{2}$O$_{5}$ coating applied
to a silica substrate. Shown are the loss factors for frequencies up
to 1kHz.}
\label{fig:newphisum}
\end{figure}

However, the thermoelastic losses for other combinations of mirror
and coating materials are estimated to be considerably larger than
is the case for SiO$_2$/Ta$_{2}$O$_{5}$ coatings on silica
substrates, see for example figure~\ref{fig:newforcompexp1}, curve
(b) for an Al$_2$O$_3$/Ta$_{2}$O$_{5}$ coating on a silica
substrate. Thus this form of dissipation should be considered in
the interpretation of measurements of coating loss factors. Since
this form of dissipation is frequency dependent, it is clearly
important to estimate the magnitude of the dissipation in the
frequency range of interest for gravitational wave detectors.
Figure~\ref{fig:newphisum} plots the same loss factors shown in the
figures above, focussing on frequencies up to approximately 1kHz.

  From figure~\ref{fig:newphisum} it can be seen that the
thermoelastic loss in the gravitational-wave detection band is
lower than at the higher frequencies sampled by mechanical loss
measurements, however it can still be of the order of $10^{-4}$.
It should be noted that in the absence of dissipation from other
sources thermal noise arising from coating-related thermoelastic
losses will form a limit to the thermal-noise performance of
interferometric detectors in a manner similar to the thermoelastic
noise from the substrates themselves.~\cite{Braginsky}
Section~\ref{sec:ThermalNoise} thus addresses the derivation of an
expression for the thermal noise from coatings arising from
thermoelastic dissipation, numerical results from which are
presented in section~\ref{subsec:numthermalnoise}.

\subsection{Thermal noise}\label{subsec:numthermalnoise}
\begin{figure}[!ht]
\includegraphics[width=9.5cm]{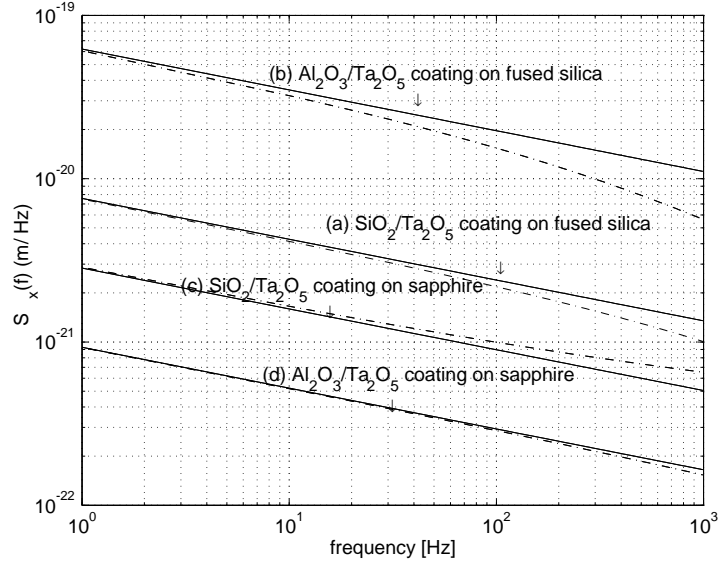}
\caption{Calculated thermoelastic thermal noise, $\sqrt{S_{x}(f)}$
for coatings of (a) an
SiO$_2$/Ta$_{2}$O$_{5}$ coating applied
to a silica substrate (b) an Al$_2$O$_3$/Ta$_{2}$O$_{5}$ applied
to a silica substrate (c) an
SiO$_2$/Ta$_{2}$O$_{5}$ coating applied
to a sapphire substrate and (d) an
Al$_2$O$_3$/Ta$_{2}$O$_{5}$ coating applied
to a silica substrate. Shown in each case is the noise obtained using the
explicit low frequency limit for the noise (solid line), and the
noise including the full frequency dependence (dashed line), as
given in Eq.~\ref{eq:Sxavgsum}. } \label{fig:newthsum}
\end{figure}

Using Eqs.~\ref{eq:Sxavgsum} and~\ref{eq:noiseresultavg} with the
parameters for coating and substrate properties given earlier, the
thermal displacement noise resulting from thermoelastic
dissipation, $\sqrt{S_{x}(f)}$, associated with silica and
sapphire mirrors with coatings of either SiO$_2$/Ta$_2$O$_5$ or
Al$_2$O$_3$/Ta$_2$O$_5$ can be estimated. Here, multilayer
coatings of a thickness equivalent to 10 ppm transmission were
modelled, since this represents a typical specification for the
transmission of a mirror coating used in the Fabry-Perot arm
cavities of a gravitational wave detector.
Figure~\ref{fig:newthsum} shows the calculated noise for each
case, for frequencies up to 1 kHz. A beam radius, $w$, of 5.5 cm
was assumed.

For comparison, the target level for total displacement noise per
test mass in the Advanced LIGO gravitational wave interferometer
design is approximately 6 x 10$^{-21} $m/$\sqrt{\text{Hz}}$ at 100
Hz if sapphire mirrors are used and approximately  8 x 10
$^{-21}$m/$\sqrt{\text{Hz}}$ for silica
mirrors.~\cite{AdvLIGOdesigndoc} Figure~\ref{fig:newthsum} shows
that for SiO$_2$/Ta$_{2}$O$_{5}$ coatings on silica or sapphire
substrates and Al$_2$O$_3$/Ta$_{2}$O$_{5}$ coatings on sapphire
substrates, the expected coating-related thermoelastic
displacement noise is below the required specification at 100 Hz.
However for an Al$_2$O$_3$/Ta$_{2}$O$_{5}$ coating on a fused
silica substrate the noise from this dissipation mechanism alone
is above the specification at 100 Hz. It is also clear that the
same coating will result in a different level of noise if applied
to different substrates.

There are several other points illustrated by
figure~\ref{fig:newthsum} worth consideration. Firstly, for some
coating/substrate combinations the thermoelastic noise starts to
deviate significantly from the explicit low frequency limit in the
frequency range of interest for detector operation. Thus when
estimating the expected level of this noise it is important to use
the full frequency-dependent expression. In addition, comparing
figures~\ref{fig:newphisum} and~\ref{fig:newthsum} it can be seen
that whilst the thermoelastic loss for strain fields associated
with typical loss measurements, $\phi_{l, \, \parallel}$ is higher
for a SiO$_2$/Ta$_{2}$O$_{5}$ coating on a sapphire substrate than
for that coating on a silica substrate, the opposite trend holds
for the thermoelastic displacement noise sensed in an
interferometer. This seeming contradiction can be understood by
comparing Eqs. \ref{eq:answeruni} for $\phi_{l, \,
\parallel}$ and \ref{eq:phiuniperp} for $\phi_{l, \, \perp}$, the loss for a specified surface-normal stress. We
see that the dependence of these two loss coefficients on the
material properties is quite different, in particular containing a ratio of
Young's moduli in the former but not the latter. In fact, for the
material properties characteristic of these coatings, unlike
$\phi_{l, \, \parallel}$, $\phi_{l, \, \perp}$ follows the same
trend as $S_x$, consistent with the observation that the axial
stress is large in the region of high optical intensity for the
fields of Eq. \ref{eq:eps0r} used in the noise calculation.

It is also important to note that in each case the exact level of
thermoelastic noise is a strong function of certain of the
material parameters of the coatings, in particular the coefficient
of thermal expansion, and given the lack of information available
on the thermoelastic properties of ion-beam-sputtered coatings,
our calculations here were carried out using plausible rather than
definitive values for relevant coating material parameters. Thus
these figures should be taken as estimates of the expected
thermoelastic noise due to the coatings, rather than reliable
results that can be used in design calculations.

\section{Detailed Calculation} \label{sec:DetailedCalc}
\subsection{Thermal fields} \label{sec:Thermal fields}
We start by calculating the thermal field $\theta(z,t)$ generated by
the applied zeroth-order elastic fields from Appendix
\ref{ap:zeroth}. We can cast the one-dimensional heat equation in the
form~\cite{LifshitzandR}
\begin{equation}
\frac{{\partial \theta _j }}{{\partial t}} - \kappa _j
\frac{{\partial ^2 \theta _j }}{{\partial z^2 }} =  - \frac{{E_j
\alpha _j T}}{{(1 - 2\nu _j )C_j }}\frac{\partial }{{\partial
t}}\sum\limits_{i = 1}^3 {\varepsilon _{0,ii,j} }
\label{eq:heateq}
\end{equation}
where $\theta_j$ is the time-varying temperature,
$\kappa_j=k_{j}/C_j$ is the thermal diffusivity, $E_j$ is the
Young's modulus, $\nu_j$ is the Poisson ratio, $T$ is the background
temperature, $C_j$ is the heat capacity per unit volume,
$\varepsilon_{0,ii,j}$ is the zeroth-order $i$-polarized
compressional strain, and $j=f,s$ indicates quantities evaluated in
the film and the substrate, respectively. Taking sinusoidally time
varying quantities according to $a(z,t)={\mathrm
{Re}}[a(z)\exp(i\omega t)]$,
Eq.~\ref{eq:heateq} becomes
\begin{equation}
i\omega \theta _j (z) - \kappa _j \frac{{\partial ^2 \theta _j
(z)}}{{\partial z^2 }} =  - i\omega \beta _j \label{eq:2.3}
\end{equation}
where the source term is proportional to
\begin{equation}
\beta _j  \equiv \frac{{E_j \alpha _j T}}{{C_j }}\frac{{\Sigma_j
}}{{1 - 2\nu _j }} \;\; . \label{eq:betaf}
\end{equation}
The quantity $\Sigma_j$ representing the sum of the strains according to
\begin{equation}
\Sigma_j \equiv \sum\limits_{i = 1}^3 {\varepsilon _{0,ii,j} }  \, ,
\label{eq:sdef}
\end{equation}
is proportional to the zeroth-order elastic field's amplitude with
a combination of elastic constants that depends on the specific
case, and is evaluated in Appendix \ref{ap:zeroth}. $\alpha_f$ is
possibly $z$-dependent to allow for spatially varying thermal
expansion coefficient within the film. The boundary conditions are
zero heat flux at $z = 0$, continuity of heat flux at $z = l$, and
vanishing heat flux for $z \to \infty$, or
\begin{equation}
\left. {\frac{{d\theta _f }}{{dz}}} \right|_{z = 0}  =
0,\,\,\,k_{f} \left. {\frac{{d\theta _f }}{{dz}}} \right|_{z =
l}  = k_{s} \left. {\frac{{d\theta _s }}{{dz}}} \right|_{z = l}
,\,\,\ \left. {\frac{{d\theta _s }}{{dz}}} \right|_{z \to
\infty}=0 \; , \label{eq:2.5}
\end{equation}
respectively.

The total solution can be constructed as the sum of homogeneous and
particular solutions. Homogeneous solutions meeting the boundary
conditions at $z = 0$ and $z \to \infty$ are
\begin{equation}
\theta _{h,f} (z) = \theta _{1f} \, \cosh (\gamma_f \,z)
\mbox{$\mathrm{\, and \ }$} \theta _{h,s} (z) = \theta _{1s}  e^{ -
\gamma _s z} \label{eq:2.6}
\end{equation}
where the complex propagation constants of the damped thermal waves
in the film and substrate are
\begin{equation}
\gamma _j  \equiv (1 + i)\sqrt {\omega /2\kappa _j } \label{eq:gamma}
\end{equation}
and $\theta _{1f} $ and $\theta _{1s}$ are constants determined by
the boundary condition at $z = l$ and the particular solution.

The particular solutions will be evaluated for two specific cases of
practical interest in Appendix \ref{ap:theta}. For the time being
take them to be $\theta _{p,j}(z)$, so that the total solutions are
\begin{eqnarray}
\theta _f (z)& =& \theta _{p,f} (z) + \theta _{1f} \cosh (\gamma _f
\,z) \, \ , \nonumber \\
\theta _s (z) &=& \theta _{p,s}  + \theta _{1s} e^{ - \gamma _s z} \,
\label{eq:2.9}
\end{eqnarray}
Note that both $\theta _{p,j} (z)$ and $\theta _{1,j}$ will be
proportional to the amplitude of the zeroth-order elastic fields.

\subsection{Elastic fields and energy
density}\label{sec:EnergyDensity}
The rate at which work is done per unit volume by internal stresses on a
deformed
body is~\cite{Landau}
\begin{equation}
p = \sigma _{ik} \frac{{d\varepsilon _{ik} }}{{dt}} \,\,.
\end{equation}
This expression is correct independent of whether the body responds
elastically or anelastically to the stresses. The cycle average of
the delivered power density (or, equivalently, the average dissipated
power density)
is then, for fields sinusoidal in time
of the form
$\sigma_{ii,j}(z,t)=\textrm{Re}\{[\sigma_{0,ii,j}+\sigma_{1,ii,j}(z)]\exp(i
\omega t)\}$ and similar for $\varepsilon_{ii,j}(z,t)$,
\begin{eqnarray}
p_{diss} (z) &=& -\frac{\omega }{2}\sum\limits_{i = 1}^3 {\mbox{Im}
\left[\sigma _{ik}^* \varepsilon _{ik} \right] } \nonumber  \\
       &=& -\frac{\omega }{2}\sum\limits_{i = 1}^3
{\mbox{Im}\left[{\sigma _{0,ii}^* \varepsilon _{0,ii}  + \sigma
_{0,ii}^* \varepsilon _{1,ii}  + \sigma _{1,ii}^* \varepsilon
_{0,ii}^{} + \sigma _{1,ii}^* \varepsilon _{1,ii} }\right]} \nonumber
\\
&\approx& -\frac{\omega }{2}\sum\limits_{i = 1}^3 {\mbox{Im}\left[{
\sigma _{0,ii}^* \varepsilon _{1,ii}  + \sigma _{1,ii}^* \varepsilon
_{0,ii}^{} }\right]} \,\, . \label{eq:p_diss_def}
\end{eqnarray}
The last form of  this equation is justified by the following
observations: For our problem, only the longitudinal strains are
significant and the zeroth-order elastic fields are real (Appendix
\ref{ap:zeroth}, and Eqs. \ref{eq:eps0r}), while the first-order
elastic fields (those that depend on the thermal field, Appendix
\ref{ap:first}) are complex and so contribute to the dissipation.
We assume that the dissipation is small, so that a calculation to
lowest order in the thermal field will be adequate, and second
order terms can be dropped.

To evaluate Eq. \ref{eq:p_diss_def}, we need the zeroth- and
first-order elastic fields in the film and the substrate, as
derived in Appendices \ref{ap:zeroth} and \ref{ap:first}. It is
convenient to write the zeroth-order fields in terms of $\sigma_0$
and $\varepsilon_0$, two combinations of the fields that are
invariant through the region of interest in the body. These are
defined by $\sigma_0 \equiv \sigma_{0,zz}$, the compressional
stress normal to the surface of the object, and $\varepsilon_0
\equiv (\varepsilon_{0,xx}+\varepsilon_{0,yy})/2$, the in-plane
dilation. For convenience, we can take $\varepsilon_{0,xx} =
\varepsilon_{0,yy}$, though only their sum matters for the
thermoelastic calculation. The antisymmetric combination of
in-plane strains is a pure shear, does not interact
thermoelastically, and so be can neglected in this analysis (as is
discussed in section \ref{sec:unifilm}). Note that in cylindrical
coordinates $\varepsilon_0 =
(\varepsilon_{0,rr}+\varepsilon_{0,\theta\theta})/2$. The
zeroth-order fields can then be summarized as
\begin{equation}\label{eq:A0B0def}
     \varepsilon _{0,ii,j}  = A_{0,ii,j} \varepsilon _0 + a_{0,ii,j}
\sigma _0 ,\,\,\sigma _{0,ii,j}  = B_{0,ii,j} \varepsilon _0 +
b_{0,ii,j} \sigma _0 \,\, ,
\end{equation}
where $j=f,s$, and the ($A_{0,ii,j}$ and $B_{0,ii,j}$) and the
($a_{0,ii,j}$ and $b_{0,ii,j}$) are combinations of elastic
constants given in Eqs. \ref {eq:A0B0} and \ref{eq:sfA0B0},
respectively. Similarly, it is convenient to write the first-order
elastic fields as proportional to the local temperature and
thermal expansion coefficient, since we assume that the
frequencies of interest are low enough that the elastic response
can be treated quasistatically:
\begin{equation} \label{eq:A1B1def}
\varepsilon _{1,ii,j} (z) = A_{1,ii,j} \alpha _j \theta _j
(z),\,\,\sigma _{1,ii,j} (z) = B_{1,ii,j} \alpha _j \theta _j (z)
\end{equation}
The (real) coefficients $A_{1,ii,j}$ and $B_{1,ii,j}$ are given in Eqs.
\ref{eq:A1B1}.

With Eqs. \ref{eq:A0B0def} and \ref{eq:A1B1def} in Eq.
\ref{eq:p_diss_def}, the dissipated power density can be written as
\begin{equation}
p_{\mathrm{diss},j}(z)=\omega \frac{\alpha _j }{2} (D_j
\varepsilon _0 +d_j \sigma_0 )   {\mathop{\rm Im}\nolimits}
[-\theta _j (z)]
\end{equation}
where
\begin{eqnarray}
D_j \equiv \sum\limits_{i = 1}^3 {\left[ {B_{0,ii,j} A_{1,ii,j}  -
B_{1,ii,j} A_{0,ii,j} } \right]}  \nonumber \\
d_j \equiv \sum\limits_{i = 1}^3 {\left[ {b_{0,ii,j} A_{1,ii,j} -
B_{1,ii,j} a_{0,ii,j} } \right]} \ . \label{eq:Ddef}
\end{eqnarray}
The dissipated power per unit area is then obtained by integrating
the power density over the thickness of the body,
\begin{eqnarray} \label{eq:PdissAB}
\frac{P_{\mathrm{diss}}}{{\rm area}} &=& \int_0^\infty
{p_{\mathrm{diss}} (z)\,dz}
\nonumber \\
     &=& \frac{\omega }{2}\left\{ {(D_f \varepsilon _0  + d_f \sigma _0
)\int_0^l {\alpha _f (z){\mathop{\rm Im}\nolimits} [ - \theta _f (z)}
]\,dz} \right. \nonumber \\
     &+& \,\,\,\,\,\,\,\, \left. (D_s \varepsilon _0  + d_s \sigma _0
)\alpha _s \int_l^\infty  {{\mathop{\rm Im}\nolimits} [ - \theta _s
(z)] \,dz} \right\}
\end{eqnarray}
where we allow for the possibility of spatial variation in the
thermal expansion coefficient in the film but assume that it is
uniform in the substrate.

We can use this expression both to evaluate the dissipation that
would be measured in a typical measurement of the $Q$ of a coated test mass
and to calculate the coating-related thermoelastic contribution to the
displacement noise imposed
on an optical field incident on a test mass. The most convenient form
of the analysis is somewhat different in these two contexts. We begin
with the case of a $Q$ measurement.

\subsection{Effective thermoelastic losses in measurements of elastic
Q} \label{sec:elasticQ}
While the total dissipated power given in Eq. \ref{eq:PdissAB} is the
physical quantity of importance to measurements of $Q$, and is
nonlocal in nature, occurring in both the film and the substrate, it
generally occurs in a region thin compared to the dimensions of the
test mass, so for comparison with experimental results it is
convenient to describe the loss in terms of an effective $\phi$
associated with the coating. To define such an effective $\phi$, we
must compare the dissipated power to some stored energy. A reasonable
choice of stored energy for the definition of $\phi$ is that in the
film, i.e.
\begin{eqnarray}
      U_{\mathrm{stor}}/{\rm area} &=&\frac{l }{2}\sum\limits_{i = 1}^3
{\mbox{Re}
\left[\sigma _{0,ii,f}^* \varepsilon _{0,ii,f} \right] } \nonumber \\
&=&l\frac{{\left| {\varepsilon _0^2 } \right|}}{2}U_f,
\label{eq:U_storeAB}
\end{eqnarray}
where
\begin{equation}
U_f \equiv \sum\limits_{i = 1}^3 {B_{0,ii,f} A_{0,ii,f} } \,\, .
\label{eq:Fdef}
\end{equation}
In writing Eqs. \ref{eq:U_storeAB} and \ref{eq:Fdef} we assumed
that the film is on a stress-free surface, so that $\sigma_{0,zz}$
and hence $\sigma_0$ vanish. $U_f$ is calculated in Appendix
\ref{ap:zeroth}, Eq. \ref{eq:Feq}. We then have for $\phi$, with
Eq. \ref{eq:PdissAB}
\begin{eqnarray}
\phi &=& \frac{{P_{\mathrm{diss}} \tau }}{{2\pi U_{\mathrm{stor}} }}
\nonumber \\
&=& \phi_f + \phi_s
\end{eqnarray}
where
\begin{equation}
\phi_f \equiv \frac {D_f}{U_f l} \int_0^l {{ \alpha _f (z)
\mathop{\rm Im}\nolimits} [-\theta _f (z)/\varepsilon _0 ]} \, dz
\label{eq:phifdef}
\end{equation}
and
\begin{equation}
\phi_s \equiv \frac {D_s}{U_f l}  \alpha _s \int_l^\infty {{
\mathop{\rm Im}\nolimits} [-\theta _s (z)/\varepsilon _0 ]} \, dz
\,\, . \label{eq:phisdef}
\end{equation}
Note that since the thermal fields are proportional to
$\varepsilon_0$, the quantity in square brackets in Eqs.
\ref{eq:phifdef} and \ref{eq:phisdef} is independent of
$\varepsilon_0$, as are all the other factors in these equations.

To make further progress, we must find the particular solution and
the coefficients for the homogeneous solutions for the thermal
field in the specific cases of interest. We consider the specific
cases of a uniform film on a uniform substrate, and a periodic
film on a uniform substrate. In Appendix \ref{ap:theta} we obtain
the thermal fields for these two cases.

\subsubsection{Uniform film and substrate} \label{sec:unifilm}
Consider first the simple model of a uniform film on a substrate,
with possibly different thermophysical properties in film and
substrate. For this case, we can take the thermal expansion
coefficients in the film and substrate to be $\alpha_f (z)
=\alpha_f$ and $\alpha_s$, respectively, and the particular
solutions to the heat equation, given in Appendix \ref{ap:theta}
as Eq. \ref{eq:thetapuni}, are
\begin{equation}
\theta _{p,j} (z) =  - \beta _j \, .
\end{equation}
Since the particular solutions are real, the only contribution to
the imaginary part of the integrals in Eqs. \ref{eq:phifdef} and
\ref{eq:phisdef} come from the homogeneous solutions, so with Eqs.
\ref{eq:2.6} we have for the film,
\begin{eqnarray}
\phi_f &=& \frac {D_f}{U_f l \,} \int_0^l {{ \alpha
_f (z) \mathop{\rm Im}\nolimits} [-\theta _f
(z)/\varepsilon _0 ]} \, dz \nonumber \\
     &=& \frac {D_f \alpha_f}{U_f l } {\mathop{\rm Im}\nolimits}
\left[  -(\theta_{1,f}/ \varepsilon_0)\int_0^l {
\cosh(\gamma_f z)\, dz} \right]  \nonumber \\
&=& \frac {D_f \alpha_f}{U_f l } {\mathop{\rm Im}\nolimits} \left[
-(\theta_{1,f}/ \varepsilon_0) \gamma_f^{-1} \sinh(\gamma_f l)
\right] \label{eq:phifuni}
\end{eqnarray}
and similarly for the substrate
\begin{equation}
\phi_s = \frac{D_s \alpha _s}{U_f l } {\mathop{\rm
Im}\nolimits} \left[ -(\theta_{1,s} /\varepsilon_0) \gamma_s^{-1}
e^{-\gamma_s l}
\right] \label{eq:phisuni} \, .
\end{equation}
where the $D_j$ are combinations of elastic constants defined in
Eq. \ref{eq:Ddef} and calculated in Eq. \ref{eq:dj}. Summing the
contributions to the dissipation from the film and the substrate,
Eqs. \ref{eq:phifuni} and \ref{eq:phisuni}, we can express the
total loss as
\begin{equation} \label{eq:phiuni}
\phi_{l,\, \parallel}=  \frac {D_f \alpha _f}{U_f l
} {\mathop{\rm Im}\nolimits} \left[ -(\theta_{1,f}/\varepsilon_0)
\gamma_f^{-1} \sinh(\gamma_f l) \right] + \frac {D_s \alpha
_s}{U_f l} {\mathop{\rm Im}\nolimits}
\left[-(\theta_{1,s}/\varepsilon_0) \gamma_s^{-1} e^{-\gamma_s l}
\right] \,\, ,
\end{equation}
where the subscript $l$ is used to indicate a quantity resulting
from thermoelastic behavior over the entire thickness of the film
(contrasted to multilayer case in following section) and
$\parallel$ indicates the case of specified in-plane strain (in
contrast to $\perp$ for specified surface-normal stress). With the
coefficients $\theta_{1,j}$ from Appendix \ref{ap:theta} Eqs.
\ref{eq:theta1funi} and \ref{eq:theta1suni} inserted into Eq.
\ref{eq:phiuni}, we can express the total loss, after some
algebra, as
\begin{eqnarray}
\phi_{l,\parallel}&=& \frac{\Delta \beta}{\varepsilon_0}
\frac{C_f(1-\nu_f)}{E_f}\left[\frac {\alpha_f E_f}{1-\nu_f} -
\frac{\alpha_s E_s}{1-\nu_s}
\frac{R \gamma_f}{\gamma_s} \right] g(\omega) \nonumber \\
&=& \frac{2 C_f (1-\nu_f) T}{E_f} \left[\frac{\alpha_f E_f}{C_f
(1-\nu_f)} - \frac{\alpha_s
E_s}{C_s (1-\nu_s)} \right]^{2} g(\omega) \nonumber \\
&=& \frac{2 E_f \alpha_f^2 T}{C_f(1-\nu_f)}\left[ 1 -
\frac{\alpha_s}{\alpha_f} \frac{E_s (1-\nu_f)}{E_f
(1-\nu_s)}\frac{C_f}{C_s} \right]^2 g(\omega) \label{eq:answeruni}
\end{eqnarray}
where the frequency dependence is contained in the function
$g(\omega)$ defined by
\begin{equation} \label{eq:gofgamma}
g(\omega)\equiv {\mathop{\rm Im}\nolimits} \left[ - \frac
{1}{\sqrt{i\omega\tau_f}} \frac{\sinh
(\sqrt{i\omega\tau_f})}{\cosh(\sqrt{i\omega\tau_f})+R\sinh(\sqrt{i\omega\tau_f})}\right]
\,\, .
\end{equation}
In deriving this result, we made use of $D_j$ from Eq.
\ref{eq:dj}, $U_f$ from Eq. \ref{eq:Feq}, $R$ from Eq.
\ref{eq:theta1s}, $\Delta \beta \equiv \beta_f - \beta_s$ from Eq.
\ref{eq:betaf}, $\Sigma$ from Eqs. \ref{eq:sigmadef} and
\ref{eq:s}, and defined $\tau_f \equiv l^2/\kappa_f$ so that with
Eq. \ref{eq:gamma} we have $\gamma_f l = \sqrt{i \omega \tau_f}$.
Note that the combination of material properties in square
brackets in Eq. \ref{eq:answeruni} is positive definite and
vanishes if the film and substrate properties are identical. The
quadratic dependence on the {\it difference} between substrate and
film properties can lead to dissipation that is sensitive to small
changes in the film properties.

A similar analysis can be carried out for an antisymmetric
in-plane strain, $\varepsilon_{0,xx}=-\varepsilon_{0,yy}$. We find
that $\Sigma_j = 0$, so that no thermal wave is generated
(consistent with the observation that this antisymmetric strain is
a pure shear along axes rotated $\pi/4$ with respect to $x$ and
$y$, causes no volume change, and hence does not contribute to the
source term for the thermal wave). We also find that $D_j=0$,
indicating that there will be no power dissipated by the
interaction of the zeroth-order antisymmetric strain with the
first-order strain fields generated by the thermal wave (driven by
other zeroth-order strains possibly present). This latter
observation can be explained by noting that the thermal wave
generates no first-order shear strains (for the geometry
considered here), and that there is no energy term associated with
the product of shear and compressional strains in isotropic media.

While not encountered in the measurement of elastic loss in coated
masses, an expression for the dissipation for a specified
surface-normal stress is useful for developing an understanding of
the thermal noise results in comparison to results for loss
measurements. Following the same analysis as was used to find
$\phi_{l, \, \parallel}$, but replacing the stored energy in Eq.
\ref{eq:U_storeAB} with
\begin{equation} \label{eq:U_storeab}
      U_{\mathrm{stor}}/{\mathrm{area}} = l\frac{{\left| {\sigma _0^2 }
\right|}}{2}u_f,
\end{equation}
where
\begin{equation}
u_f \equiv \sum\limits_{i = 1}^3 {b_{0,ii,f} a_{0,ii,f} } \,\, .
\label{eq:fdef}
\end{equation}
we find that replacing $U_f \rightarrow u_f$, $D_j \rightarrow
d_j$, and $\varepsilon_0 \rightarrow \sigma_0$ in Eq.
\ref{eq:phiuni} yields the correct result for $\phi_{l, \,
\perp}$,
\begin{equation} \label{eq:phiuniperp}
\phi_{l,\, \perp}=\frac{E_f \alpha_f^2
T}{C_f}\frac{1+\nu_f}{(1-\nu_f)(1-2\nu_f)}\left[ 1 -
\frac{\alpha_s}{\alpha_f}
\frac{(1-\nu_f)(1+\nu_s)}{(1-\nu_s)(1+\nu_f)}\frac{C_f}{C_s}
\right]^2 g(\omega) \,\, ,
\end{equation}
where we made use of $d_j$ from Eq. \ref{eq:sfdj}, $u_f$ from Eq.
\ref{eq:sfFeq}, $\Sigma$ from Eqs. \ref{eq:sigmadef} and
\ref{eq:sfs}, other quantities as after Eq. \ref{eq:gofgamma}, and
$g(\omega)$ is the same frequency dependence given in Eq.
\ref{eq:gofgamma}.

\subsubsection{Modulated film and uniform substrate} \label{sec:modfilm}
Optical coatings of interest for use on test masses for
gravitational wave detectors are invariably multilayered, so the
analysis in the previous section of a uniform film cannot be
correct in detail. Because the thermal diffusion length for
frequencies of interest are in general long compared to the
thickness of individual layers in the film, an accurate
approximation for a multilayer coating can be obtained by using a
suitable averaging process to model it as a uniform layer (except
at very high frequencies), as discussed in Appendix \ref{ap:avg}.
Following the procedure described there, we find that the result
in Eq.~\ref{eq:answeruni} is replaced by
\begin{eqnarray} \label{eq:phiuniavg}
\phi_{l, \, \parallel}= \frac{2 C_F T}{\left( \frac{E}{1-\nu}
\right)_{\mathrm{avg}}} \left[  \frac{1}{C_F}\left(\frac{E
\alpha}{1-\nu} \right)_{\mathrm{avg}}-\frac{1}{C_s}\frac{E_s
\alpha_s}{1-\nu_s} \right]^2 g(\omega)
\end{eqnarray}
and the frequency dependence $g(\omega)$ is unchanged except for
replacing the time constant $\tau_f$ by an appropriately averaged
one $\tau_F$. The volume-weighted average indicated by
$(X)_{\mathrm{avg}}$ is defined in Eq.~\ref{eq:avgdef}, the
averaged heat capacity $C_F$ in Eq.~\ref{eq:avgheat}, and $\tau_F$
in Eq.~\ref{eq:tauF}.

While the results based on this averaging process appear
reasonable, it is useful to explore for unexpected effects
associated with the spatial variation of thermoelastic properties
within the multilayer film. As a simple model of such a case, we
consider a film whose thermal expansion coefficient, but no other
property, varies periodically, on a uniform substrate. The
calculation is similar in principal, but somewhat more complicated
than for the case of a uniform film on a uniform substrate. For
this case, we take a thermal expansion coefficient of the form
\begin{equation}
\alpha_f (z) = \overline{\alpha}_f + \alpha_m \cos(K_m z) \mbox{ and
} \alpha_s(z) = \alpha_s \label{eq:alphamod}
\end{equation}
and the particular solution to the heat equation in the film, given
in Appendix \ref{ap:theta} as Eq. \ref{eq:thetapfmod} is
\begin{equation}
\theta _{p,f} (z) =  - \beta _f  - \beta _m \frac{{\gamma
_f^2 }}{{\gamma _f^2  + K_m^2 }}\cos (K_m z) \,
\label{eq:thetapfmodmain}
\end{equation}
while the particular solution in the substrate, Eq.
\ref{eq:thetapsmod}, remains the same as for the uniform film case
\begin{equation}
\theta _{p,s}^{}  =  - \beta _s \, . \label{eq:thetapsmodmain}
\end{equation}

For simplicity, we consider here only the case of specified
in-plane strain and vanishing surface-normal stress $\sigma_0$.
Since the particular solution in the film has a complex part, the
imaginary part of the integrals in Eqs. \ref{eq:phifdef} and
\ref{eq:phisdef} will contain contributions from both the
particular and the homogeneous solutions. With Eqs. \ref{eq:2.6},
\ref{eq:alphamod} and \ref{eq:thetapfmodmain} we have in the film
\begin{eqnarray}
\phi_f &=& \frac {D_f}{U_f l} \int_0^l {{ \alpha _f (z)
\mathop{\rm Im}\nolimits} [-\theta _f
(z)/\varepsilon _0 ]} \, dz \nonumber \\
&=& \frac {D_f} {U_f l}(\Phi_{f,p} + \Phi_{f,h}) \, ,
\label{eq:phifbreakmod}
\end{eqnarray}
where
\begin{eqnarray}
\Phi_{f,p} &\equiv&  {\mathop{\rm Im}\nolimits}  \left[ -\int_0^l {
\left[ \overline{\alpha} _f+\alpha_m \cos(K_m z) \right]
[\theta_{p,f}(z)/\varepsilon_0] \, dz} \right]  \nonumber \\
&=&  \frac{\beta_m}{K_m \, \varepsilon_0}\left\{
\overline{\alpha}_f \sin(K_m l) + \alpha_m \left[ \frac{K_m
l}{2}+\frac{\sin(2K_ml)}{4} \right] \right\} {\mathop{\rm
Im}\nolimits}\left[ \frac{\gamma_f^2 }{\gamma
_f^2  + K_m^2 }\right] \nonumber \\
\label{eq:Phifpmod}
\end{eqnarray}
and
\begin{eqnarray}
\Phi_{f,h} &\equiv&  {\mathop{\rm Im}\nolimits}  \left[ -\int_0^l {
\left[ \overline{\alpha} _f+\alpha_m \cos(K_m z) \right]
[\theta_{h,f}(z) / \varepsilon_0] \, dz} \right]  \nonumber \\
&=& -{\mathop{\rm Im}\nolimits}\left[ \overline{\alpha}_f
\frac{\theta_{1,f}}{\varepsilon_0} \gamma_f^{-1} \sinh{\gamma_f l}
\right. \nonumber \\
&\,\,\,\,\,& + \left. \alpha_m \frac{\theta_{1,f}}{\varepsilon_0}
\frac{K_m \cosh(\gamma_f l)\sin(K_m l)+\gamma_f\sinh(\gamma_f
l)\cos(K_m l)}{K_m^2 + \gamma_f^2} \right] \label{eq:Phifhmod}
\end{eqnarray}

Since the thermal expansion coefficient in the substrate is assumed
uniform, the result for $\phi_s$ is similar to that of Eq.
\ref{eq:phisuni} (though of course $\theta_{1,s}$ will be different
in the two cases). We use Eqs. \ref{eq:2.6}, \ref{eq:alphamod} and
\ref{eq:thetapsmodmain} to obtain for the substrate
\begin{equation}
\phi_s = \frac {D_s \alpha _s}{U_f l} {\mathop{\rm Im}\nolimits}
\left[ -(\theta_{1,s}/\varepsilon_0) \gamma_s^{-1} e^{-\gamma_s l}
\right] \, . \label{eq:messymod}
\end{equation}
Considerably more effort is required to convert Eqs.
\ref{eq:Phifpmod}, \ref{eq:Phifhmod}, and \ref{eq:messymod} into a
simple form like Eq. \ref{eq:answeruni} after inserting
$\theta_{1,f}$ and $\theta_{1,s}$ from Eqs. \ref{eq:theta1fmod} and
\ref{eq:theta1smod}. The following two
terms emerge as dominant in Eqs. \ref{eq:Phifpmod} and
\ref{eq:Phifhmod} for cases where $|K_m|>>|\gamma_f|$,
\begin{eqnarray}
\Phi_{f,p}&\approx& \frac{\alpha_m l}{2}
\frac{\beta_m}{\varepsilon_0} \, \mbox{Im} \left[
\frac{\gamma_f^2}{\gamma_f^2+K_m^2} \right] \nonumber \\
\Phi_{f,h} &\approx& \overline{\alpha}_f \frac{\Delta
{\beta}}{\varepsilon_0} \, g(\omega) \, ,
\end{eqnarray}
so that with Eq. \ref{eq:phifbreakmod} we have
\begin{eqnarray}
\phi_f \approx \frac {D_f} {U_f l} \left\{\frac{\alpha_m l}{2}
\frac{\beta_m}{\varepsilon_0} \, \mbox{Im} \left[
\frac{\gamma_f^2}{\gamma_f^2+K_m^2} \right] + \overline{\alpha}_f
\frac{\Delta {\beta}}{\varepsilon_0} \, g(\omega)  \right\}\,\, ,
\label{eq:phifapprox}
\end{eqnarray}
where $g(\omega)$ is defined in Eq. \ref{eq:gofgamma}. Similarly,
keeping the leading term in Eq. \ref{eq:messymod} after inserting Eq.
\ref{eq:theta1smod} leads to
\begin{equation}
\phi_s \approx -\frac{\Delta {\beta}}{\varepsilon_0} \frac {D_s
\alpha _s}{U_f l}R\, g(\omega)   \label{eq:phisapprox}
\end{equation}
where all neglected terms are smaller by at least one factor of
$\gamma_f / K_m$ than those retained. The first term in Eq.
\ref{eq:phifapprox} is unique to a modulated film. The second term
in Eq. \ref{eq:phifapprox} and Eq. \ref{eq:phisapprox} are just
the same as those that appeared in the expression for a
homogeneous film and substrate, Eq. \ref{eq:phiuni} with $\alpha_f
\rightarrow \overline{\alpha}_f$, so that portion of the solution
can be used here immediately. Rewriting Eqs. \ref{eq:phifapprox}
and \ref{eq:phisapprox} with $D_j$ from Eq. \ref{eq:dj}, $U_f$
from Eq. \ref{eq:Feq}, $R$ from Eq. \ref{eq:theta1s}, $\beta_m$
from Eq. \ref{eq:betam}, $\beta_j$ from Eq. \ref{eq:betaf},
$\gamma_j$ from Eq. \ref{eq:gamma}, and $\Sigma_j$ from Eq.
\ref{eq:sigmadef} with Eq. \ref{eq:s} we obtain
\begin{equation} \label{eq:phimod}
\phi_{\mathrm{tot}, \, \parallel} \approx
\frac{E_fT\alpha_m^2}{C_f(1-\nu_f)}\mbox{Im} \left[
\frac{\gamma_f^2}{\gamma_f^2+K_m^2} \right] + \phi_{l, \,
\parallel} \, ,
\end{equation}
where $\phi_{l, \, \parallel}$ is the loss for a uniform film on a
uniform substrate given in Eq. \ref{eq:answeruni}, with $\alpha_f
\rightarrow \overline{\alpha}_f$. Defining a characteristic time
for diffusion in the modulated structure,
\begin{equation}\label{eq:tau_m}
       \tau_m \equiv \frac{\tau_f}{K_m^2l^2}=\frac{1}{\kappa_f K_m^2}
\end{equation}
where the second form follows from the definition of $\tau_f$
following Eq. \ref{eq:gofgamma}, $\phi_{\mathrm{tot}, \,
\parallel}$ can be written in the form
\begin{equation}
        \label{eq:phimod2}
\phi_{\mathrm{tot},\, \parallel} \approx
\frac{E_fT\alpha_m^2}{C_f(1-\nu_f)} g_m(\omega) + \phi_{l, \,
\parallel} \, ,
\end{equation}
where the frequency dependence $g_m(\omega)$ is
\begin{equation} \label{eq:gmofomega}
       g_m(\omega)=\frac{\omega \tau_m}{1+\omega^2 \tau_m^2} \,\, .
\end{equation}

In section \ref{sec:FrequencyDependence} we show that for the
frequencies of interest, the frequency dependence $g_m(\omega)$
from Eq. \ref{eq:gmofomega} representing heat flow between the
multilayers is, as expected, small compared to $g(\omega)$ from
Eq. \ref{eq:gofgamma} representing heat flow between film and
substrate, so that considering only the contribution of the
averaged form $\phi_{l, \,\parallel}$ from Eq. \ref{eq:phiuniavg}
is a good approximation.

\subsubsection{Frequency dependence} \label{sec:FrequencyDependence}
The functions $g(\omega)$ and $g_m(\omega)$, defined in Eqs.
\ref{eq:gofgamma} and \ref{eq:gmofomega}, respectively, contain all the
frequency dependence of the dissipation, and will reappear in our
discussion of displacement
noise in section \ref{sec:ThermalNoise}. It is therefore worth
investigating their general features in some detail. Consider first
$g(\omega)$; it is convenient to define the real variable $\xi$
according to
\begin{equation}
\gamma_f l =\sqrt{i \omega \tau_f}\equiv (1+i) \xi/2
\end{equation}
so that
\begin{eqnarray} \label{eq:xsi}
\xi \equiv \sqrt{\frac{2\omega l^{2}}{\kappa_f}}=\sqrt{2 \omega
\tau_f} \,\,
\end{eqnarray}
where from Eq. \ref{eq:gofgamma} $\tau_f = l^2 C_f/k_f \,$. The frequency
dependence of the loss due to a uniform film can
then be written as
\begin{eqnarray} \label{eq:gofxsi}
g(\xi)= \xi^{-1}\frac{\sinh\xi-\sin\xi+R
(\cosh\xi-\cos\xi)}{\cosh\xi+\cos\xi+2 R
\sinh\xi+R^{2}(\cosh\xi-\cos\xi)}\,\, .
\end{eqnarray}

\begin{figure}[ht!]
\includegraphics[width=12cm]{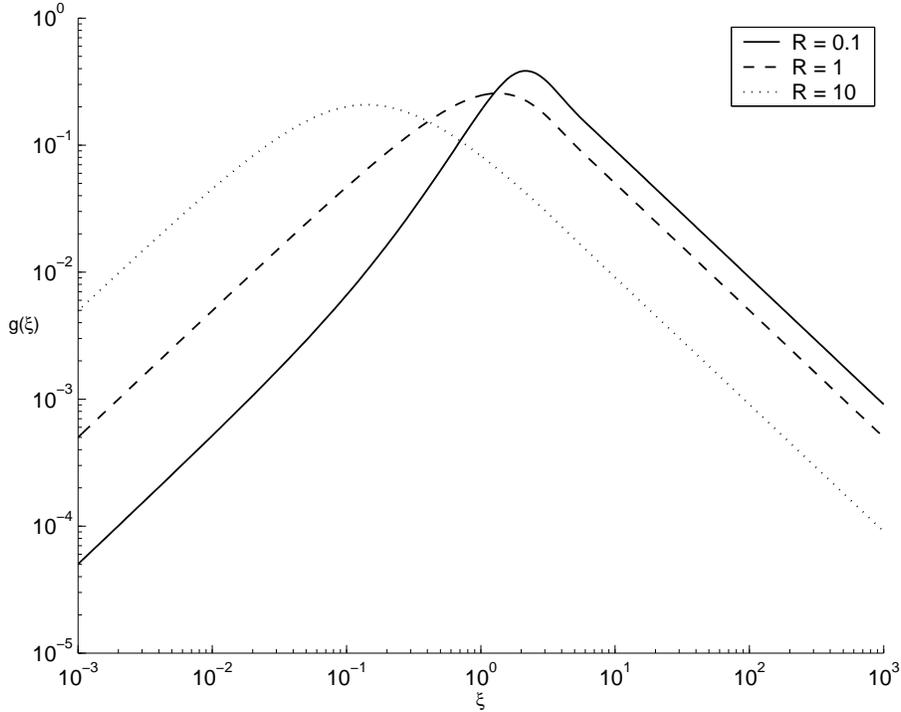}
\caption{Frequency dependent part of thermoelastic loss function,
$g(\xi)$ as a function of frequency, where $\xi \equiv \sqrt{2 \omega
\tau_f}$ and $\tau_f = l^{2} C_{f}/k_{f}$.
Curves are shown for three values of the parameter $R \equiv k_{f}
\gamma _f /k_{s} \gamma _s = \sqrt{k_f C_f/k_s C_s}$. Note that
peak height is only a weak function of $R$. }
\label{fig:phivsxi}
\end{figure}

\begin{figure}[ht!]
\includegraphics[width=12cm]{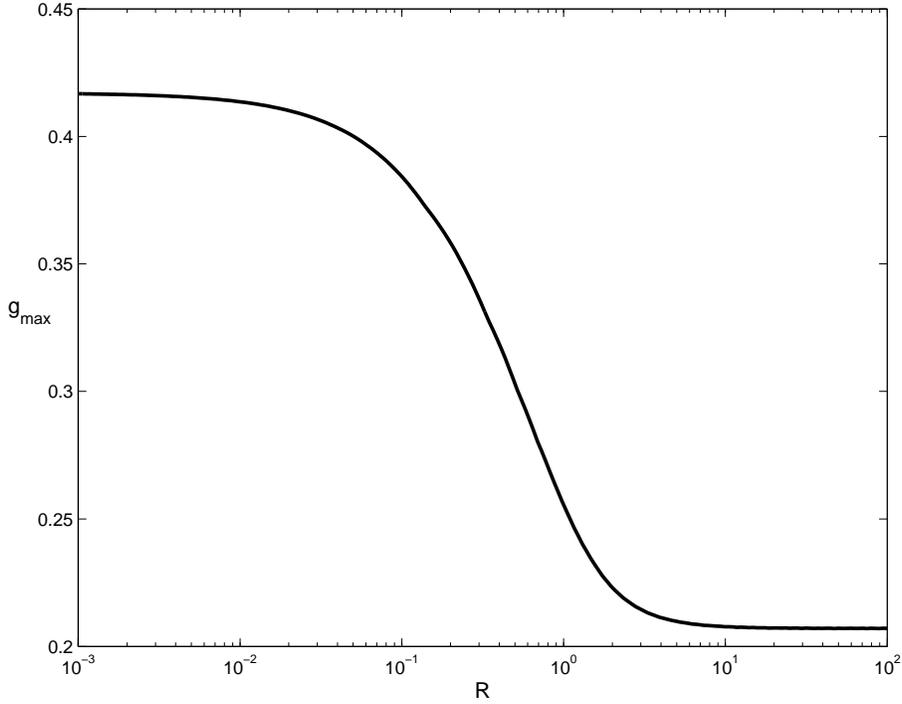}
\caption{Peak value of the normalized dissipation,
$g_{\mathrm{max}}\equiv g(\xi_{\mathrm{max}})$ as a function of
the parameter $R$. The peak value of the normalized dissipation is
seen to be only a weak function of $R$, the only material
parameter on which it depends.} \label{fig:phimaxvsR}
\end{figure}

\begin{figure}[ht!]
\includegraphics[width=12cm]{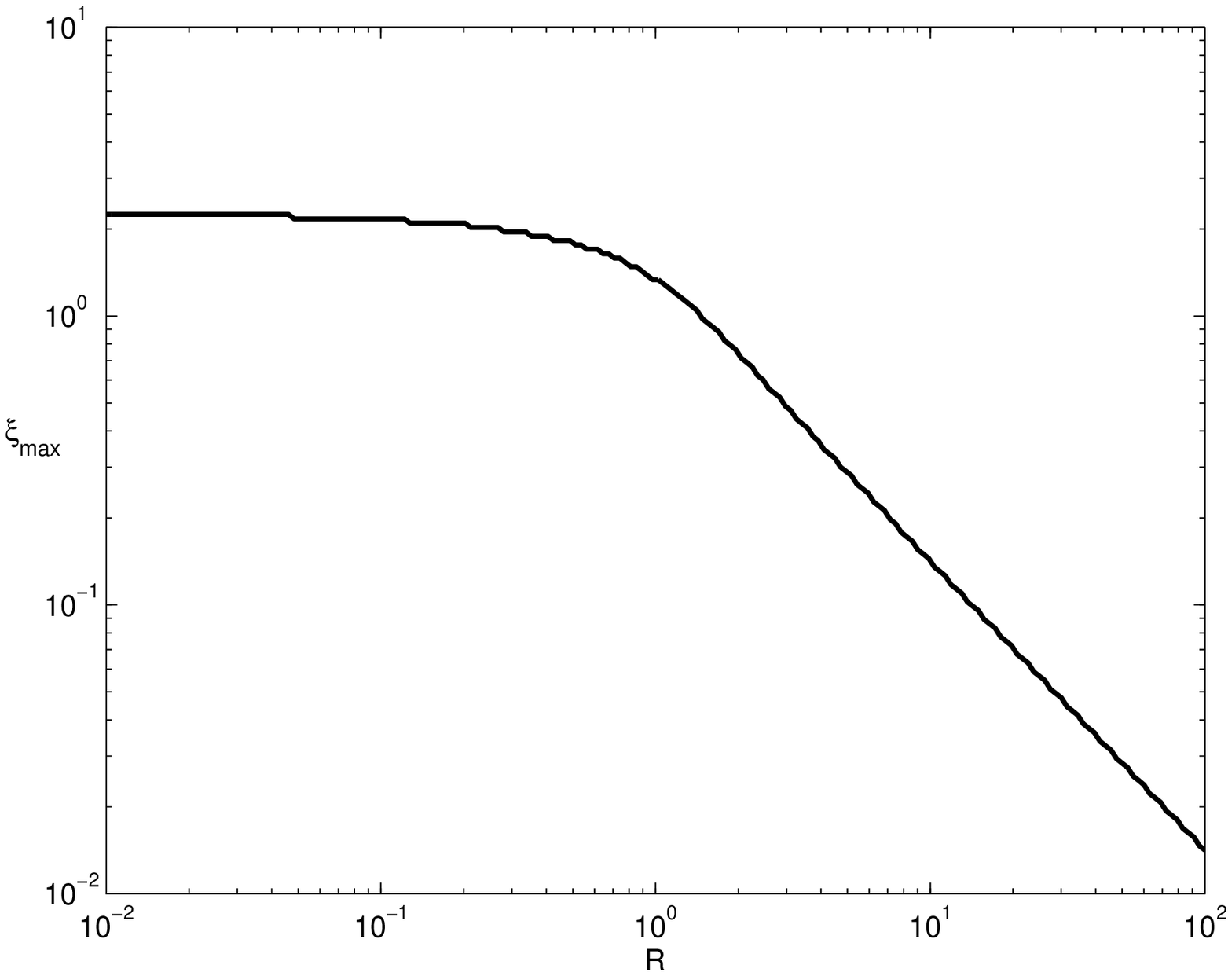}
\caption{Normalized frequency $\xi_{\mathrm{max}}$ at which the
normalized dissipation takes its maximum value, as a function of
the parameter $R$. $\xi_{\mathrm{max}}$ corresponds to the thermal
diffusion time across the film for $R<1$, and decreases  as $1/R$
(corresponding to $\omega_{\mathrm{max}}$ decreasing as $1/R^{2}$
for $R>1$). } \label{fig:xsimaxvsR}
\end{figure}
In terms of the normalized frequency $\xi$, the normalized
dissipation depends only on one parameter, $R$, defined after Eq.
\ref{eq:theta1s}  $\,$as $R \equiv k_{f} \gamma _f /k_{s} \gamma
_s = \sqrt{k_f C_f/k_s C_s}$. Figure \ref{fig:phivsxi} shows the
frequency dependence of the dissipation for $R = (0.1, 1, 10)$.
Figure \ref{fig:phimaxvsR} shows the dependence on $R$ of
$g_{\mathrm{\mathrm{max}}}$, the peak value of the normalized
dissipation, and figure \ref{fig:xsimaxvsR} the dependence on $R$
of $\xi_{\mathrm{max}}$, the normalized frequency at which this
peak occurs. We see that $g_{\mathrm{max}}$ depends only weakly on
$R$, ranging from 0.41 for $R \ll 1$ to 0.21 for $R \gg 1$.
$\xi_{\mathrm{max}}$ is close to the thermal diffusion time across
the film ($\xi = 2$) for $R \leq 1$, and decreases as $1/R$ with
$R$ for $R > 1$.

Useful forms for low- and high-frequency limits of the dissipation
can be obtained from Eq. \ref{eq:gofxsi}. Expanding for $\xi\ll 1$ we
find
\begin{equation} \label{eq:glow}
g(\xi)\rightarrow \frac{1}{2}\left[R \xi
-(R^{2}-\frac{1}{3})\xi^{2}\right]
\end{equation}
while for $\xi \gg 1$
\begin{equation}
g(\xi) \rightarrow \frac {1}{(1+R)\xi} \,\, .
\end{equation}
Since $\omega \propto \xi^{2}$, the leading behavior for low
frequencies goes as the $\sqrt{\omega}$, and, surprisingly, the sign
of the term linear in frequency depends on the value of $R$, crossing
zero for $R = 1/\sqrt{3}$. At high frequencies, the dissipation falls
off as $1/\sqrt{\omega}$.

The frequency dependence described by $g_m(\omega)$ for the
contribution from the multilayer coating is simple, and
essentially the same as that for conventional thermoelastic
damping, so it needs little further discussion. It is important to
note that for typical multilayer coatings, the characteristic time
$\tau_m$ for the multilayer effects is much shorter than $\tau_f$
for the effects of the averaged uniform layer; for a coating with
$2N$ layers,
\begin{equation} \label{eq:taufbytaum}
\tau_f/\tau_m = K_m^2l^2=4 \pi^2 N^2 \,\, .
\end{equation}
Since a typical high reflector that might be used in a LIGO
interferometer has ~40 layers, $\tau_f \sim 16000 \tau_m$, so that the
peak frequency for the contribution to the thermoelastic dissipation
from thermal diffusion between the layers will be at a frequency
$\sim 16000$ times higher, generally pushing the peak well above typical
measurement ranges. The thermoelastic effects at frequencies of
interest either for elastic Q measurements or thermal noise are thus
generally dominated by the contributions of the averaged film.

\subsection{Thermal noise}
\label{sec:ThermalNoise}
The results of section~\ref{subsec:numQ} indicate that the power
dissipated by the thermoelastic effects can be comparable or even
exceed that dissipated by the elastic loss in typical multilayer
coatings. It is then reasonable to assume that the magnitude of the
noise induced by the thermoelastic mechanism could be comparable to
that from the elastic loss, and therefore must be calculated as part
of the total noise budget for the coated mass.

Following the approach of~\cite{Levin}, the displacement noise
imposed on a Gaussian beam of normalized intensity distribution $I(r)$
\begin{equation} \label{Iofr}
I(r) = \frac{2}{{\pi w^2 }}\exp \left( {\frac{{ - 2r^2 }}{{w^2 }}}
\right)
\end{equation}
is given by
\begin{equation}\label{eq:LevinNoise}
S_x (f) df= \frac{{2k_B T}}{{\pi ^2 f^2 }}\frac{{W_{\mathrm{diss}} }}{{F_0^2
}}df
\end{equation}
where $W_{\mathrm{diss}}$ is the cycle-averaged power dissipated by a
pressure
field $\rho(r)$ oscillating at a frequency $\omega = 2\pi f$, of the
same radial distribution as the intensity and with a resultant force
$F_0$, i.e. \begin{equation}\label{eq:pressure}
\rho (r) = F_0 I(r)\cos (\omega t) \,\, .
\end{equation}
Since the radius of the Gaussian beam is much larger than the
thickness of the film or the thermal-wave decay length, we can use
the one-dimensional theory developed in previous sections of this
paper to evaluate $W_{\mathrm{diss}}$. The zeroth-order elastic fields
required for this
calculation are available in~\cite{Harry}. We have from their Eqs.
(A10)
\begin{eqnarray}\label{eq:eps0r}
     \sigma_0(r) &\equiv& \sigma_{0,zz} \nonumber \\
      &=& -\rho(r) \nonumber \\
      \varepsilon _0 (r) &\equiv& [\varepsilon _{rr} (r) + \varepsilon
_{\theta \theta } (r)]/2 \nonumber \\
       &=& -\rho (r)/4(\lambda  + \mu ) \nonumber \\
       &=& -\frac{(1+\nu_s)(1-2\nu_s)}{2E_s}\rho(r)
\end{eqnarray}
where the third form of $\varepsilon_0$ follows from the definition
of the Lam$\acute{\text{e}}$ constants in terms of the Young's modulus and
Poisson
ratio.

Here we analyze the noise due to an appropriately averaged uniform
film, since the analysis of section~\ref{sec:FrequencyDependence}
showed that the contribution to the dissipation associated with
the thermal diffusion between the layers within the film are
significant only at frequencies well above the LIGO detection
band. We first consider a uniform film, and then the modifications
necessary to describe an appropriately averaged multilayer.

Starting with Eq. \ref{eq:PdissAB} for the dissipated power per unit
area, and Eqs. \ref{eq:2.9} for the thermal fields we have
\begin{eqnarray}
      \frac{{P_{\mathrm{diss}} }}{{{\rm area}}} &=& \frac{\omega }{2}\left\{
\alpha _f (D_f \varepsilon _0  + d_f \sigma _0 ){\mathop{\rm
Im}\nolimits} [ - \theta _{1f} \gamma _f^{ - 1} \sinh (\gamma _f l)]
\right. \nonumber \\
      &+& \left.  \,\alpha _s (D_s \varepsilon _0  + d_s \sigma _0
){\mathop{\rm Im}\nolimits} [ - \theta _{1s} \gamma _s^{ - 1} \exp (
- \gamma _f l)] \right\} \nonumber \\
&=& \frac{\omega \Delta \beta l}{2} \left[ \alpha_f (D_f
\varepsilon_0+d_f \sigma_0)- \alpha_s (D_s \varepsilon_0+d_s
\sigma_0)\frac{C_f}{C_s}\right]g(\omega)
\label{eq:Pdisstherm}
\end{eqnarray}
where as in Eqs. \ref{eq:phifuni} and \ref{eq:phisuni}, we simplified
the result by noting that only the homogeneous part of the thermal
solutions contains an imaginary part, and in the second form used
Eqs. \ref{eq:theta1funi} and \ref{eq:theta1suni} for $\theta_{1,f}$
and $\theta_{1,s}$. The frequency dependence is contained in the same
function $g(\omega)$ defined in Eq. \ref{eq:gofgamma}.

For this case, where there are both axial stresses and in-plane
strains, we can calculate $\Delta \beta \equiv \beta_f - \beta_s$
from Eqs. \ref{eq:betaf}, \ref{eq:sdef}, \ref{eq:sigmadef},
\ref{eq:s}, and
\ref{eq:sfs}. We find that
\begin{eqnarray}
\beta_j &=& \frac{E_j\alpha_j T}{C_j}\frac{1}{1-\nu_j}\left[ 2
\varepsilon_0 + \frac{1+\nu_j}{E_j}\sigma_0 \right] \nonumber \\
&=& \frac{E_j\alpha_j T}{C_j}\frac{1}{1-\nu_j}\left[
\frac{(1+\nu_s)(1-2\nu_s)}{E_s} + \frac{1+\nu_j}{E_j} \right]
\sigma_0 \label{betanoise}
\end{eqnarray}
where the second follows for the specific form of the elastic fields
given in Eqs. \ref{eq:eps0r}. With Eqs. \ref{eq:dj} and \ref{eq:sfdj}
for the combinations of elastic constants represented by $D_j$ and
$d_j$, and $\Delta \beta$ calculated from Eq. \ref{betanoise}, the
result in Eq. \ref{eq:Pdisstherm} for the dissipated power per unit
area becomes
\begin{eqnarray}\label{pdissfinal}
        \frac{P_{\mathrm{diss}}(r)}{{\rm area}}&=& \frac{\omega T l
C_f}{2}\rho(r)^{2}g(\omega) \nonumber \\
&\times& \left\{\frac{\alpha_f}{C_f} \left[
\frac{1+\nu_f}{1-\nu_f}+\frac{(1+\nu_s)(1-2\nu_s)}
{1-\nu_f}\frac{E_f}{E_s}\right]-\frac{\alpha_s}{C_s}2 (1+\nu_s)
\right\}^2
\end{eqnarray}
where we replaced $\sigma_0^{2}$ with $\rho(r)^2$ according to Eq.
\ref{eq:eps0r}. Integrating over the infinite cross-section to obtain
the total dissipated power $W_{\mathrm{diss}}$, and inserting that result
into
Eq. \ref{eq:LevinNoise} for $S_x(f)$, we finally obtain
\begin{eqnarray} \label{eq:noiseresult}
S_x(f)&=& \frac{8k_BT^{2}}{\pi^{2}f}\frac{l}{w^2}C_f\,g(\omega)
\nonumber \\
&\times& \left\{\frac{\alpha_f}{C_f}\frac{1}{2}\left[
\frac{1+\nu_f}{1-\nu_f}+\frac{(1+\nu_s)(1-2\nu_s)}
{1-\nu_f}\frac{E_f}{E_s}\right]-\frac{\alpha_s}{C_s} (1+\nu_s)
\right\}^2 \nonumber \\
&=&
\frac{8k_BT^{2}}{\pi^{2}f}\frac{l}{w^2}\frac{\alpha_s^{2}C_f}{C_s^{2}}(1+\nu_s)^{2}\Delta^2
g(\omega)
\end{eqnarray}
where $\Delta^{2}$ is a dimensionless positive-definite combination
of material constants that vanish when the film and substrate are
identical,
\begin{equation} \label{eq:delta}
\Delta^{2}\equiv\left\{\frac{C_s}{2 \alpha_s
C_f}\frac{\alpha_f}{(1-\nu_f)}\left[
\frac{1+\nu_f}{1+\nu_s}+(1-2\nu_s)\frac{E_f}{E_s}\right]-1
\right\}^2 \,\, .
\end{equation}
Eq. \ref{eq:noiseresult} is the final result for the thermoelastic
displacement noise associated with a uniform coating. The frequency
dependence represented by $g(\omega)$ is discussed at length in
section \ref{sec:FrequencyDependence}.

An accurate approximation for a multilayer coating can be obtained by
using a suitable averaging process to model it as a uniform layer, as
discussed in Appendix \ref{ap:avg}. Following the procedure described
there, the result in Eq.~\ref{eq:noiseresult} is replaced by
\begin{equation} \label{eq:noiseresultavg}
S_x(f)=\frac{8k_BT^{2}}{\pi^{2}f}\frac{l}{w^2}\frac{\alpha_s^{2}C_F}{C_s^{2}}(1+\nu_s)^{2}
\tilde{\Delta}^2 g(\omega)
\end{equation}
where
\begin{equation} \label{eq:deltilde}
\tilde{\Delta}^2\equiv\left\{ \frac{C_s}{2 \alpha_s C_F} \left(
\frac{\alpha}{1-\nu}\left[
\frac{1+\nu}{1+\nu_s}+(1-2\nu_s)\frac{E}{E_s}\right]\right)_{\mathrm{avg}}-1
\right\}^2
\,\, ,
\end{equation}
and the frequency dependence $g(\omega)$ is unchanged except for
replacing the time constant $\tau_f$ by an appropriately averaged
one $\tau_F$. The volume-weighted average indicated by
$(X)_{\mathrm{avg}}$ is defined in Eq.~\ref{eq:avgdef}, the
averaged heat capacity $C_F$ in Eq.~\ref{eq:avgheat}, and $\tau_F$
in Eq.~\ref{eq:tauF}. Since for room-temperature operation the
thermoelastic noise is most important at frequencies falling in
the low-frequency limit of $g(\omega)$, it is useful to insert
into Eq.~\ref{eq:noiseresultavg} the approximate result for $g$
given in Eq. \ref{eq:glow} to obtain
\begin{equation} \label{eq:sxlow}
S_x(f) \to \frac{8 \sqrt{2}k_B T^2}{\pi
\sqrt{\omega}}\frac{l^2}{w^2}(1+\nu_s)^2 \,
\frac{C_F^2}{C_s^2}\frac{\alpha_s^2}{\sqrt{k_s C_s}} \, \tilde{\Delta}^2
\,\, .
\end{equation}
Note that, as seen in figure \ref{fig:newthsum}, this low
frequency limit becomes inaccurate at the upper end of the
gravitational-wave detection band.

\section{Summary and conclusions}
In this paper we have derived expressions for the thermoelastic
dissipation associated with a coating on a test mass. For strains
of the type consistent with mechanical loss measurements,
numerical evaluation of the thermoelastic loss factors for
coating/test mass material combinations of the type being
considered for use in future gravitational wave interferometers
shows that thermoelastic dissipation is of a level comparable to
that predicted to affect the sensitivity of advanced
interferometers. Also derived is an expression for the expected
power spectral density of thermoelastic noise from the coating of
a mirror interrogated with a Gaussian beam. Evaluating this
expression across the gravitational wave detector band using
plausible values for the material parameters of coatings and
substrates results in displacement noise that in some cases
exceeds typical design sensitivities.

It should be noted that the expected thermoelastic noise is a
strong function of the difference of the material parameters in
the substrate and coating, so that the same coating will have
different thermoelastic losses on different substrates. As many of
the necessary material parameters are not well characterized, the
noise levels calculated here should be considered as estimates
only. Further experimental measurements of coating dissipation for
likely choices for coating and substrate materials, and better
characterization of the intrinsic coating thermophysical
properties, should allow more accurate determination of the
magnitude of the thermoelastic effects.

\begin{acknowledgments}
The authors are supported by NSF grants PHY-0140297 (MF, SR),
PHY-0107417 (GH), PHY-9801158 (AG) and PHY-0140335 (SP), and
PHY-0098715 (SV). SR, DC, JH, GC and PS also thank PPARC in the UK,
and the University of
Glasgow for financial support. SV also thanks the Russian Ministry of
Industry and Science and the Russian Foundation of Basic Researches. We
also wish to thank Vladimir Braginsky for useful discussions, and our
colleagues in the GEO 600 project and at Stanford for their interest in
this work.
\end{acknowledgments}

\appendix
\section{Zeroth-order driving fields}
\label{ap:zeroth}
We need a set of zeroth order driving fields that are consistent with
the elastic boundary conditions. In all cases we must have continuity
of the in-plane strains and the normal stress at the film-substrate
interface, $z = l$: $\varepsilon _{0,xx,s}  = \varepsilon _{0,xx,f}
,\,\,\,\varepsilon_{0,yy,s}  = \varepsilon _{0,yy,f}$ and
$\sigma_{0,zz,f}= \sigma_{0,zz,s}$. As discussed in section
\ref{sec:intro}, the pertinent elastic fields can be specified in
terms of two components, the in-plane dilation $\varepsilon_{0}
\equiv (\varepsilon_{0,xx}+\varepsilon_{0,yy})/2$ and the axial
stress $\sigma_0 \equiv \sigma_{0,zz}$, which are independent of $z$
under the assumptions set up in section \ref{sec:intro}. We can
neglect the anti-symmetric in-plane strain
($\varepsilon_{0,xx}-\varepsilon_{0,yy}$) which does not interact
thermoelastically (as is shown in section \ref{sec:unifilm}), and, for
convenience, can take
$\varepsilon_{0,xx}=\varepsilon_{0,yy}=\varepsilon_{0}$.

We consider two cases, a stress-free surface with a specified
in-plane strain ($\sigma_0=0$ and $\varepsilon_0$ specified), and a
specified surface-normal stress with a vanishing in-plane
strain ($\varepsilon_0=0$ and $\sigma_0$ specified). Any elastic state
pertinent to the thermoelastic problem can be obtained as an
appropriately weighted sum of these two solutions. For the general
case, where both $\sigma_0$ and $\varepsilon_0$ are nonzero, one of
the important results of this appendix, $\Sigma_j$ defined in Eq.
\ref{eq:sdef}, can be written
\begin{eqnarray}\label{eq:sigmadef}
       \Sigma_j & \equiv & \sum\limits_{i = 1}^3 {\varepsilon_{0,ii,j} }
\nonumber \\
       & = & S_{j} \varepsilon_0+s_{j}\sigma_0 \,\, .
\end{eqnarray}
The combination of elastic constants $S_j$ and $s_j$ are obtained in
this Appendix, Eqs. \ref{eq:s} and \ref{eq:sfs}, respectively.

\subsection{Specified in-plane strain, stress-free surface}
\label{ap:stressfree}
For a stress-free surface of the mass, as would be the case for a
Q-measurement, we have $\sigma_0=0$ and $\varepsilon_0$ specified.
Noting that the under these assumptions the continuity condition on
the normal stress implies that
$\sigma_{0,zz,s}=\sigma_{0,zz,f}=\sigma_{0}= 0$, the only unknown
field components are  $\varepsilon _{0,zz,f},\, \varepsilon
_{0,zz,s},\,\sigma_{0,xx,f}=\sigma_{0,yy,f}\equiv\sigma_{0,\|,f}$,
and $\sigma_{0,\|,s}$. The symmetry of the problem allowed us to take
$\sigma _{0,xx}= \sigma_{0,yy}\equiv \sigma_{0,\|}$.

Begin with Hooke's law, Eq. 5.14 of~\cite{Landau}:
\begin{eqnarray}
      \varepsilon _{0,zz,j}  &=& \frac{1}{E_j}[\sigma _{0,zz,j}  -
\nu_{j}(\sigma _{0,xx,j}  + \sigma _{0,yy,j} )] \nonumber \\
       &=& \frac{{ - 2\nu_{j} }}{E_{j}}\sigma _{0,\|,j} \,\, .
\label{eq:A2.37}
\end{eqnarray}
where we recall the notation that a subscript $j=f,s$ stands for a
quantity evaluated in the film or substrate, respectively. Summing
the expressions for the in-plane strains in~\cite{Landau},
it follows that
\begin{equation}
\varepsilon _{0,xx,j}  + \varepsilon _{0,yy,j}  =
\frac{1}{E_j}[\sigma _{0,xx,j}  + \sigma _{0,yy,j}  - \nu_j (\sigma
_{0,xx,j}  + \sigma _{0,yy,j} )]
\end{equation}
or equivalently
\begin{equation}
\varepsilon _{0}  = \frac{{1 - \nu_j }}{E_j}\sigma _{0,\|,j} \,\, .
\label{eq:A2.39}
\end{equation}
Finally, going back to Eq.~\ref{eq:A2.37} with Eq. \ref{eq:A2.39}, we
find
\begin{equation}
\varepsilon _{0,zz,j}  = \frac{{ - 2\nu_j}}{{1 - \nu_j}}\varepsilon
_0 \,\, . \label{eq:A2.40}
\end{equation}
It is convenient to summarize these results for the zeroth-order
elastic fields in the form given in Eq. \ref{eq:A0B0def},
\begin{equation}
\varepsilon _{0,ii,j} = A_{0,ii,j} \varepsilon _0
,\,\,\,\,\,\sigma _{0,ii,j} = B_{0,ii,j} \varepsilon _0,
\label{eq:zerothfields}
\end{equation}
where
\begin{eqnarray}
     A_{0,xx,j}  = A_{0,yy,j}  = 1,\,\,A_{0,zz,j}  = \frac{{ - 2\nu _j
}}{{1 - \nu _j }} \nonumber \\
     B_{0,xx,j}  = B_{0,yy,j}  = \frac{{E_j }}{{1 - \nu _j
}},\,\,B_{0,zz,j}  = 0  \label{eq:A0B0}
\end{eqnarray}

A result, used in Eq. \ref{eq:heateq}, is the evaluation of a sum
over strains introduced in Eq. \ref{eq:sdef}, which, with Eqs.
\ref{eq:sigmadef} and \ref{eq:A0B0}, becomes
\begin{eqnarray} \label{eq:s}
     S_j \varepsilon_0 \equiv \sum\limits_{i = 1}^3 {\varepsilon _{0,ii,j}
}=\sum\limits_{i = 1}^3 {A _{0,ii,j} \varepsilon_0}=\frac{{2(1 -
2\nu_j )}}{{1 - \nu_j }} \, \varepsilon_0 \,\, .
\end{eqnarray}
Another result, used in Eq. \ref{eq:Fdef} to evaluate the energy
stored in the film is
\begin{equation}  \label{eq:Feq}
U_f \equiv \sum\limits_{i = 1}^3 {B_{0,ii,f} A_{0,ii,f} }=
\frac{{2E_f }}{{1 - \nu _f }}  \,\, .
\end{equation}
\subsection{Specified surface-normal stress, vanishing in-plane
strain}
For a specified surface-normal stress, as would be the case for
calculating thermal noise, we have $\sigma_{0,zz,f} = \sigma_0$. To
make this case complementary to that in section \ref{ap:stressfree},
we assume vanishing in-plane strains, i.e. $\varepsilon _{0,xx}  =
\varepsilon _{0,yy} \equiv \epsilon_{0}= 0$ in both the film and the
substrate. Noting
that the under these assumptions the continuity condition on the
normal stress implies that $\sigma_{0,zz,s}=\sigma_{0,zz,f} =
\sigma_0$, so the only unknown field components are  $\varepsilon
_{0,zz,j}$ and $\sigma_{0,xx,j}=\sigma_{0,yy,j}\equiv
\sigma_{0,\|,j}$, for $j=f,s$.
The analysis is similar to that in section \ref{ap:stressfree}. Begin with
Eq. 5.14 of ~\cite{Landau}:
\begin{equation}
      \varepsilon _{0,xx,j}  = \frac{1}{E_j}[\sigma _{0,xx,j}  -
\nu_{j}(\sigma _{0,yy,j}  + \sigma _{0,zz,j} )] \nonumber
\end{equation}
which can be solved with $\sigma_{0,xx,j}=\sigma_{0,yy,j} \equiv
\sigma_{0,\|,j}$ to yield
\begin{equation}
\sigma_{0,\|,j}=\frac{\nu_j}{1-\nu_j}\sigma_0 \,\, .  \label{eq:sfsigma0xx}
\end{equation}
With another of Eqs. 5.14 from~\cite{Landau}:
\begin{equation}
      \varepsilon _{0,zz,j}  = \frac{1}{E_j}[\sigma _{0,zz,j}  -
\nu_{j}(\sigma _{0,xx,j}  + \sigma _{0,yy,j} )] \nonumber
\end{equation}
and Eq. \ref{eq:sfsigma0xx} we obtain
\begin{equation}
      \varepsilon _{0,zz,j}  =
\frac{\sigma_0}{E_j}\frac{(1-2\nu_j)(1+\nu_j)}{1-\nu_j} \,\, .
\end{equation}
We can again collect the results of this section in the form given in
Eqs. \ref{eq:A0B0def},
\[
\varepsilon _{0,ii,j} = a_{0,ii,j} \sigma_0
,\,\,\,\,\,\sigma _{0,ii,j} = b_{0,ii,j} \sigma_0,
\]
\begin{eqnarray}
     a_{0,xx,j}  &=& a_{0,yy,j}  = 0,\,\,a_{0,zz,j}  = \frac{{ (1- 2\nu
_j)(1+\nu_j) }}{{1 - \nu _j }} \frac{1}{E_j} \nonumber \\
     b_{0,xx,j}  &=& b_{0,yy,j}  = \frac{\nu_j}{1 - \nu _j }
,\,\,b_{0,zz,j}  = 1 \,\, . \label{eq:sfA0B0}
\end{eqnarray}

A result, used in Eq. \ref{eq:heateq}, is the evaluation of a sum
over strains introduced in Eq. \ref{eq:sdef}, which, with Eqs.
\ref{eq:sigmadef} and \ref{eq:sfA0B0} can be written
\begin{eqnarray} \label{eq:sfs}
     s_j \,\sigma_0 \equiv \sum\limits_{i = 1}^3 {\varepsilon _{0,ii,j} } =
\sum\limits_{i = 1}^3 {a_{0,ii,j} } \, \sigma_0= \frac{{ (1- 2\nu
_j)(1+\nu_j) }}{{1 - \nu _j }} \frac{1}{E_j}
\, \sigma_0 \,\, .
\end{eqnarray}
Another result, used in Eq. \ref{eq:fdef} to evaluate the energy
stored in the film is
\begin{equation}  \label{eq:sfFeq}
u_f \equiv \sum\limits_{i = 1}^3 {b_{0,ii,f} a_{0,ii,f} }= \frac{{
(1- 2\nu _f)(1+\nu_f) }}{E_f(1 - \nu _f )}  \,\, .
\end{equation}

\section{The thermal fields for two important cases} \label{ap:theta}
The unknown coefficients in the homogeneous parts of the thermal
fields, Eqs. \ref{eq:2.6}, can be obtained from the particular
solutions, Eqs. \ref{eq:2.9}, and the boundary conditions, Eqs.
\ref{eq:2.5}. Continuity of the thermal field at $z = l$ requires
\begin{equation}
\theta _{p,f} (l) + \theta _{1f} \cosh (\gamma _f \,l) = \theta
_{p,s}  + \theta _{1s} e^{ - \gamma _s l}
\end{equation}
while continuity of the thermal flux requires
\begin{equation}
k_f \left[ \theta '_{p,f} (l) + \theta _{1f} \gamma _f \sinh (\gamma _f
\,l) \right]
=  -k_s \, \theta _{1s} \gamma _s e^{ - \gamma _s l}.
\end{equation}
Simultaneous solution of these equations yields
\begin{equation}
\theta _{1,f}  = \frac{{\left[ \theta _{p,s}  - \theta _{p,f}(l)
\right] -
(R/\gamma _f )\theta '_{p,f} (l)}}{{\cosh (\gamma _f l) + R\sinh
(\gamma _f l)}} \label{eq:theta1f}
\end{equation}
and
\begin{equation}
\theta _{1,s}  = -e^{\gamma _s l} \frac{\left[ \theta _{p,s}  -
\theta_{p,f} (l) \right] R  \sinh ( \gamma _f l)+ (R/\gamma _f )\cosh
(\gamma _f l)\theta '_{p,f} (l)}{{\cosh (\gamma _f l) + R \sinh
(\gamma _f l)}}    \,\, \label{eq:theta1s} .
\end{equation}
where $R \equiv k_{f} \gamma _f /k_{s} \gamma _s $. To make
further progress, it is necessary to find the particular solutions
for specific cases. We consider here two cases of interest, a uniform
film on a uniform substrate, and a periodic film on a uniform
substrate.
\subsection{Uniform film on uniform substrate}
Consider first both the film and substrate to be uniform. By
inspection of the thermal field equation, Eq. \ref{eq:2.3},
particular solutions for this case are constant and given by
\begin{equation}
\theta _{p,j} (z) =  - \beta _j \,\, . \label{eq:thetapuni}
\end{equation}
With Eq. \ref{eq:thetapuni} for the particular solutions, the
coefficients in the homogeneous solutions in film and substrate from
Eqs. \ref{eq:theta1f} and \ref{eq:theta1s} become
\begin{equation}
\theta _{1,f}  = \frac{\Delta \beta}{{\cosh (\gamma _f l) + R\sinh
(\gamma _f l)}} \label{eq:theta1funi}
\end{equation}
and
\begin{equation}
\theta _{1,s}  = -e^{\gamma _s l} \frac{\Delta \beta R \sinh(\gamma_f
l)}{{\cosh (\gamma _f l) + R \sinh (\gamma _f l)}}    \,\,
\label{eq:theta1suni} .
\end{equation}
where $\Delta \beta \equiv \beta_f - \beta_s$.
\subsection{Modulated film on uniform substrate}
In the case of a nonuniform film, the expression for the particular
solution is somewhat more complicated. Assume a film whose thermal
conductivity takes the form
\begin{equation}
\alpha _f (z) = \bar \alpha _f  + \alpha _m \cos (K_m z),
\end{equation}
in which case with Eq. \ref{eq:betaf} $\beta_f(z)$ takes the form
\begin{equation}
\beta _f (z) =  \beta _f  + \beta _m \cos (K_m z)
\label{eq:betamod}
\end{equation}
where
\begin{equation} \label{eq:betam}
\beta _f  \equiv \frac{{E_f \bar \alpha
_f T}}{{C_f }}\frac{{\Sigma_f }}{{1 - 2\nu _f }} \,\,{\mathrm { and
}}\,\,\beta _m  \equiv \frac{{E_f \alpha _m T}}{{C_f
}}\frac{{\Sigma_f}}{{1 -
2\nu _f }} \,\,.
\end{equation}

With Eq. \ref{eq:betamod}, the thermal field equation Eq.
\ref{eq:2.3} takes the form
\begin{eqnarray}
      \frac{{\partial ^2 \theta _{p,f} (z)}}{{\partial z^2 }} &=&
\frac{{i\omega }}{{\kappa _f }}[\theta _{p,f} (z) + \beta _f  +
\beta _m \cos (K_m z)]\nonumber \\
       &=& \gamma _f^2 [\theta _{p,f} (z) +  \beta _f  + \beta _m
\cos (K_m z)] \label{eq:de_theta_pf}
\end{eqnarray}
where the definition $\gamma _f^2  = i\omega /\kappa _f$ from
Eq.~\ref{eq:gamma} was used to obtain the second form. The particular
solution has two terms, a constant part similar to that in Eq.
\ref{eq:thetapuni} for the uniform case, and one that has a spatial
variation that follows the thermal expansion coefficient. To obtain
the periodic part, take an ansatz $\theta _{p,f} (z) = q\,\cos (K_m
z)$. With this ansatz in Eq. \ref{eq:de_theta_pf} we obtain
\begin{equation}
(\gamma _f^2  + K_m^2 )\,q =  - \gamma _f^2 \,\beta _m
\end{equation}
Combining with the constant part we obtain the total particular
solution in the film,
\begin{equation}
\theta _{p,f} (z) =  -  \beta _f  - \beta _m \frac{{\gamma
_f^2 }}{{\gamma _f^2  + K_m^2 }}\cos (K_m z) . \label{eq:thetapfmod}
\end{equation}
For the assumed uniform substrate, the particular solution is like
that in Eq. \ref{eq:thetapuni}, i.e.
\begin{equation}
\theta _{p,s}^{}  =  - \beta _s \,\, . \label{eq:thetapsmod}
\end{equation}
With Eqs. \ref{eq:thetapfmod} and \ref{eq:thetapsmod} for the
particular solutions, the coefficients in the homogeneous solutions
in film and substrate from Eqs. \ref{eq:theta1f} and \ref{eq:theta1s}
become
\begin{equation}
\theta _{1,f}  = \frac{\Delta \beta}{{\cosh(\gamma _f l) + R
\sinh(\gamma _f l)}} + \beta_m  \frac{\gamma_f^2}{\gamma_f^2+K_m^2}
\frac{\cos(K_m l)-(R K_m/\gamma_f)\sin(K_m l)}{{\cosh (\gamma _f l) +
R\sinh(\gamma _f l)}} \label{eq:theta1fmod}
\end{equation}
and
\begin{eqnarray}
\theta _{1,s}  = &-& \Delta \beta e^{\gamma _s l} \frac{R
\sinh(\gamma_f l)}{{\cosh(\gamma _f l) + R \sinh(\gamma _f l)}}
\nonumber \\
&-& \beta_m R e^{\gamma _s l} \frac{\gamma_f^2}{\gamma_f^2+K_m^2}
\frac{\cos(K_m l)\sinh(\gamma_f l)-(K_m/\gamma_f)\sin(K_m l)\cosh
(\gamma _f l)}{{\cosh (\gamma _f l) + R\sinh(\gamma _f l)}}
\label{eq:theta1smod} \, . \nonumber \\
{}
\end{eqnarray}
where $\Delta \beta \equiv {\beta}_f - \beta_s$.

\section{Solving for the thermoelastically generated elastic fields}
\label{ap:first}
Given the solution Eqs. \ref{eq:2.9} for the oscillatory thermal
field, we must solve for the thermally driven elastic fields,
$\sigma_{1}(z)$ and $\varepsilon_{1}(z)$, whose imaginary parts lead to the
dissipation in which we are interested. The boundary conditions are
$\sigma_{1,zz}=0$ at the stress-free surface $z=0$,
$\varepsilon_{1,xx}=\varepsilon_{1,yy}=0$ for $z \rightarrow \infty$,
and continuity of the in-plane strains $\varepsilon_{1,xx}$ and
$\varepsilon_{1,yy}$, and the normal stress $\sigma_{1,zz}$, at the
boundary between the film and substrate.

The point of departure is the equation of elastic equilibrium,
Eq. 7.8 of \cite{Landau},
\begin{equation}\label{eq:thermo}
\frac{d}{{dz}}\left[ {\varepsilon _{xx}  + \varepsilon _{yy}  + 2(1 -
\nu )\varepsilon _{zz}  - 2(1 + \nu )\alpha \theta }
\right] = 0
\end{equation}
adapted here by dividing Landau's $\alpha$ by 3 to convert it from
volumetric to linear expansion, replacing $ \alpha \nabla \theta $
with $ \nabla (\alpha \theta)$ to accommodate a possible spatial
variation in the thermal expansion coefficient, and specializing to
stress and strain fields that depend only on $ z $. Hooke's law in
the presence of a nonuniform temperature field $\theta(z)$, Eq. 6.2 of
~\cite{Landau}
is:
\begin{equation}\label{eq:hookeslaw}
\sigma _{zz}  = \frac{E}{{1 + \nu }}\left[ {\varepsilon _{zz}  +
\frac{\nu }{{1 - 2\nu }}(\varepsilon _{xx}  + \varepsilon _{yy}  +
\varepsilon _{zz} )} \right] - \frac{{E\alpha \theta }}{{1 - 2\nu }}
\,\, .
\end{equation}
With the boundary condition $\sigma_{1,zz} = 0$, Eq.
\ref{eq:hookeslaw} results in
\begin{equation} \label{eq:eps1zz}
\varepsilon _{1,zz,j}  =  - \frac{{\nu _j }}{{1 - \nu _j
}}(\varepsilon _{1,xx,j}  + \varepsilon _{1,yy,j} ) + \frac{{1 + \nu
_j }}{{1 - \nu _j }}\alpha _j \theta _j \,\, .
\end{equation}
Inserting Eq. \ref{eq:eps1zz} into Eq. \ref{eq:thermo}, we find
\[ \frac{d}{dz}[(1-2\nu_j)(\varepsilon _{1,xx,j}  + \varepsilon
_{1,yy,j} )]=0 \,\, . \]
Noting the continuity of the in-plane strains, and their vanishing at
infinity, we conclude $(\varepsilon _{1,xx,j}  + \varepsilon
_{1,yy,j} )=0$. With this result, Eq. \ref{eq:eps1zz} becomes
\begin{equation} \label{eq:eps1zzfinal}
\varepsilon _{1,zz,j}  =  \frac{{1 + \nu _j }}{{1 - \nu _j }}\alpha
_j \theta _j \,\, .
\end{equation}
With the Hooke's law expression for $\sigma_{xx}$ analogous to
Eq.~\ref{eq:hookeslaw} for $\sigma_{zz}$, and inserting
Eq.~\ref{eq:eps1zzfinal}, we obtain
\begin{equation}\label{sig1xx}
        \sigma_{1,xx,j}=-\frac {E_j \alpha_j \theta_j}{1-\nu_j}
\end{equation}
and by symmetry $\sigma_{1,yy,j}= \sigma_{1,xx,j}$.

These results constitute a consistent set of first-order elastic
fields. It is convenient to summarize them in the form:
\begin{equation} \label{eq:firstfields}
\varepsilon _{1,ii,j} (z) = A_{1,ii,j} \alpha_j \theta_j
,\,\,\,\,\,\sigma _{1,ii,j} (z) = B_{1,ii,j} \alpha_j \theta_j,
\end{equation}
where
\begin{eqnarray} \label{eq:A1B1}
A_{1,xx,j}&=&A_{1,yy,j}  = 0,\,\,A_{1,zz,j}=\frac{{1 + \nu _j }}{{1 -
\nu _j }} \nonumber \\
B_{1,xx,j}&=&B_{1,yy,j}  = -\frac{{E_j }}{{1 - \nu _j
}},\,\,B_{1,zz,j}=0 \,\, .
\end{eqnarray}

Combinations of these parameters used in calculating the dissipated
power, $D_j$ and $d_j$ in Eqs. \ref{eq:Ddef}, can be evaluated with
Eqs. \ref{eq:A0B0} and \ref{eq:A1B1} as
\begin{equation}  \label{eq:dj}
D_j \equiv \sum\limits_{i = 1}^3 {\left[ {B_{0,ii,j} A_{1,ii,j}  -
B_{1,ii,j} A_{0,ii,j} } \right]} = \frac {2E_j}{1-\nu_j}  \,\, ,
\end{equation}
and with Eqs. \ref{eq:sfA0B0} and \ref{eq:A1B1} as
\begin{equation}  \label{eq:sfdj}
d_j \equiv \sum\limits_{i = 1}^3 {\left[ {b_{0,ii,j} A_{1,ii,j} -
B_{1,ii,j} a_{0,ii,j} } \right]} = \frac {1+\nu_j}{1-\nu_j} \,\, .
\end{equation}

\section{Averaging material properties in a periodic multilayer}
\label{ap:avg}
In cases of practical interest, the optical coating is a multilayer rather
than a homogeneous film. The analysis in sections \ref{sec:modfilm} and
\ref{sec:FrequencyDependence} indicated that for realistic cases either
for Q measurements or for thermoelastic noise, the thermal diffusion
length is large compared to the period of the multilayer. Hence, an
analysis that treats the film as an effective homogeneous medium with
suitably averaged properties should yield a result of adequate accuracy.
It is then necessary to form the appropriate average of the various
material properties involved. For simplicity, we take the coating to
consist of alternating layers of two types of material, labelled $a$ and
$b$, of thicknesses $d_a$ and $d_b$, respectively. We define the volume
averaging operator by
\begin{equation} \label{eq:avgdef}
(X)_{\mathrm{avg}}\equiv \frac{d_a}{d_a+d_b} X_a + \frac{d_b}{d_a+d_b} X_b
\,\, .
\end{equation}

\subsection{Thermal field averaging} \label{ap:thermalavg}
Consider first the heat equation. Define an average temperature in the
film, $\theta_F(z)$, where we use the subscript $F$ to indicate a quantity
in the film suitably averaged over a period of the structure (averaging
will be different for different quantities), under the assumption that the
distance over which this averaged quantity varies significantly is much
greater than the period $d_a + d_b$. Since the temperature field is
continuous at the boundaries between the layers,
$\theta_F(z)=\theta_a(z)=\theta_b(z)$. To obtain an averaged heat equation
for the propagation of $\theta_F$, begin with Eq. \ref{eq:2.3}, here
rewritten in a more convenient form,
\begin{equation} \label{eq:distribheat}
i\omega C_q \, \theta _q (z) -\frac{\partial}{\partial z}\left(k_q
\frac{\partial \, \theta_q}{\partial z} \right) =  - i\omega C_q \, \beta
_q \,\, ,
\end{equation}
%\begin{equation}
%C_q \, \frac{{\partial \theta _q }}{{\partial t}} -
%\frac{\partial}{\partial z}\left(k_q \frac{\partial \,
%\theta_q}{\partial z} \right) =  - \frac{{E_q \alpha _q T}}{{(1 -
%2\nu _q)}}\frac{\partial }{{\partial
%t}}\sum\limits_{i = 1}^3 {\varepsilon _{0,ii,q} } \, ,
%\end{equation}
where $q=a,b$ indicates a quantity evaluated in layer $a$ or $b$,
respectively. Averaging the first and last terms over a period of the
structure is trivial. The second term requires more care. Noting that the
continuity of the heat flux requires that $k_a \, \partial \theta_a /
\partial z = k_b \, \partial \theta_b / \partial z = (k \, \partial
\theta/ \partial z)_{\mathrm{avg}}$, implicitly defining the averaged
thermal conductivity $k_F$ by writing the average heat flux as
\[
k_F \left( \frac{\partial \theta_F}{\partial z}
\right)_{\mathrm{avg}}\equiv \left( k \, \frac{\partial \theta}{ \partial
z} \right)_{\mathrm{avg}} \,\, ,
\]
and solving for $k_F$, we find
\begin{equation} \label{eq:kavg}
k_F^{-1}=(k^{-1})_{\mathrm{avg}}\,\, .
\end{equation}
We can then write the averaged Eq. \ref{eq:distribheat} in a form
analogous to Eq. \ref{eq:2.3}, \begin{equation} \label{eq:avgheat}
i\omega \, \theta _F (z) -\kappa_F\frac{\partial^2 \theta_F}{\partial z^2}
=  - i\omega \beta _F \,\, ,
\end{equation}
where the averaged film properties are
\begin{eqnarray} \label{eq:thermavg}
C_F &\equiv& (C)_{\mathrm{avg}}  \nonumber \\
\kappa_F &\equiv& k_F/C_F  \nonumber \\
\beta_F &\equiv& \frac{(C \, \beta)_{\mathrm{avg}}}{C_F} =\frac{1}{C_F}
\left( \frac{E \alpha T \Sigma}{1 - 2\nu}\right)_{\mathrm{avg}}
\end{eqnarray}
where we used Eq.~\ref{eq:betaf} for $\beta$. With these averaged
quantities in place of those of the uniform film, i.e. taking $X_f \to
X_F$, we can immediately transcribe all the previous results for the
temperature field in the uniform film without further analysis. It is also
convenient to define a thermal diffusion time for the averaged film of
thickness $l$,
\begin{equation} \label{eq:tauF}
\tau_F \equiv l^2/\kappa_F \,\, .
\end{equation}

\subsection{Elastic fields in a multilayer} \label{ap:zerothavg}
Averaging of the elastic properties is more straightforward. The
zeroth-order elastic fields already taken as invariant through the
region of interest, i.e. the in-plane dilation $ \varepsilon_0
\equiv (\varepsilon_{0,xx} +\varepsilon_{0,yy})/2 $ and the axial
stress $\sigma_0 \equiv \sigma_{0,zz}$, remain invariant in the
multilayer, so they are obviously equal to their average.

The calculation of the remaining components of the elastic field then
follows exactly as given in appendix \ref{ap:zeroth}, so that the correct
result for the fields in material $q=a,\, b$ in a modulated film can be
obtained from the corresponding expression for the zeroth-order field in a
uniform film $j=f$ by replacing $j \to q$. For example, for the case
$\varepsilon_0$ specified and $\sigma_0 = 0$ we simply have, analogously
to Eq. \ref{eq:zerothfields}
\begin{equation}  \label{eq:zerothqfields}
\varepsilon _{0,ii,q} = A_{0,ii,q} \varepsilon _0
,\,\,\,\,\,\sigma _{0,ii,q} = B_{0,ii,q} \varepsilon _0,
\end{equation}
where
\begin{eqnarray} \label{eq:A0B0q}
     A_{0,xx,q}  = A_{0,yy,q}  = 1,\,\,A_{0,zz,q}  = \frac{{ - 2\nu _q
}}{{1 - \nu _q }} \nonumber \\
     B_{0,xx,q}  = B_{0,yy,q}  = \frac{{E_q }}{{1 - \nu _q}},\,\,B_{0,zz,q}
= 0  \,\, .
\end{eqnarray}
The same approach provides the results for $a_{0,ii,q}, \, b_{0,ii,q}, \,
\Sigma_{0,ii,q}, \, S_q, \, s_q$, from the corrsponding expressions in
Eqs. \ref{eq:sfA0B0}, \ref{eq:sigmadef}, \ref{eq:s}, and \ref{eq:sfs},
respectively.

For the power stored in the film, analogous to Eq.~\ref{eq:Feq}, we must
average the energy stored in the components of the multilayer,
\begin{equation} \label{eq:Feqq}
U_F = \left( \frac{2E}{1 - \nu} \right)_{\mathrm{avg}} \,\,.
\end{equation}

By similar arguments as were applied to the zeroth-order fields,
the first-order fields analogous to those obtained for a uniform
film in Appendix \ref{ap:first}, can be obtained from the
corresponding expression for the first-order field in a uniform
film $j=f$ by replacing $j \to q$. For example, analogously to
Eqs. \ref{eq:firstfields},
\begin{equation} \label{eq:firstfieldsq}
\varepsilon _{1,ii,q} (z) = A_{1,ii,q} \alpha_q \theta_F(z)
,\,\,\,\,\,\sigma _{1,ii,q} (z) = B_{1,ii,q} \alpha_q \theta_F(z),
\end{equation}
where
\begin{eqnarray} \label{eq:A1B1q}
A_{1,xx,q}&=&A_{1,yy,q}  = 0,\,\,A_{1,zz,q}=\frac{{1 + \nu _q }}{{1 -
\nu _q }} \nonumber \\
B_{1,xx,q}&=&B_{1,yy,q}  = -\frac{{E_q }}{{1 - \nu _q
}},\,\,B_{1,zz,q}=0 \,\, .
\end{eqnarray}
The combinations of these parameters used in calculating the dissipated
power, $D_q$ and $d_q$, analogous to Eqs.~\ref{eq:dj} and \ref{eq:sfdj},
are obtained similarly.

\subsection{Averaging the dissipated power} \label{ap:poweravg}
To find the averaged dissipated power, start with Eq. \ref{eq:PdissAB}.
Noting that the temperature is continuous and slowly varying over a period
of the structure, we can write
\begin{eqnarray} \label{eq:PdissABavg}
\frac{P_{diss}}{{\rm area}}&=& \frac{\omega }{2}\left\{ \left( (D
\varepsilon _0  + d \sigma _0) \alpha\right)_{\mathrm{avg}} {\int_0^l
{{\mathop{\rm Im}\nolimits} [ - \theta _F (z)}
]\,dz} \right. \nonumber \\
     &+& \,\,\,\,\,\,\,\, \left. (D_s \varepsilon _0  + d_s \sigma _0
)\alpha _s \int_l^\infty  {{\mathop{\rm Im}\nolimits} [ - \theta _s
(z)] \,dz} \right\} \,\, .
\end{eqnarray}
Note that in making this approximation, we exclude cases where the thermal
diffusion length approaches the layer period, but do allow the thermal
length to be less than to the total thickness of the multilayer. This is
not a very restrictive assumption: for typical mirror films of $\sim 20$
layer pairs, frequencies up to $\sim 10^4$ above the dissipation peak are
allowed (see Eq. \ref{eq:taufbytaum}). Comparing with
Eq.~\ref{eq:PdissAB}, we see that any result for a uniform film can be
transformed into the corresponding result for the averaged film by
replacing
\begin{equation} \label{eq:pdisscorr}
    (D_f \varepsilon _0  + d_f \sigma _0) \alpha_f \to  \left( (D \varepsilon
_0  + d \sigma _0) \alpha\right)_{\mathrm{avg}} \,\, ,
\end{equation}
where the elastic quantities required are given in
Eqs.~\ref{eq:dj} and \ref{eq:sfdj}, and the averaging operation is
defined in Eq.~\ref{eq:avgdef}; we again replace the thermal
properties with the averaged ones given in Eqs.~\ref{eq:kavg} and
\ref{eq:thermavg}, i.e. $X_f \to X_F$. Since the dissipated power
is the key quantity from which all the end results of this paper
devolve, only straightforward substitution and algebraic
manipulation are required to obtain those results for the averaged
film. The results so obtained for the dissipation factor $\phi$
and spectral density of thermoelastic noise $S_x(f)$ are given in
Eqs.~\ref{eq:phiuniavg} and \ref{eq:noiseresultavg}, respectively.


\end{document}